\documentclass[twocolumn]{raa_twocolumn}            % referee version: for submission

\usepackage{graphicx,times}             %for PS/EPS graphics inclusion, new
\usepackage{natbib}
\usepackage{amssymb,amsmath}
\pdfoutput=1
\bibpunct{(}{)}{;}{a}{}{,}

\usepackage[pagebackref=true]{hyperref}

\begin{document}

  \title{The radial velocity stability for reference sources in the MWISP CO survey}
%   \subtitle{I. Place Your Subtitle Here}

   \volnopage{Vol.0 (202x) No.0, 000--000}      %%preserved for Editor. DOn't remove!
   \setcounter{page}{1}          %%starting page, preserved for Editor. DOn't remove!

   \author{Lixia Yuan %% Put your Chinese name in "( )" if you like. 
      \inst{1}\footnotetext{$*$}\footnotetext{ID: https:/orcid.org/0000-0003-0804-9055}
   \and Ji Yang
      \inst{1}\footnotetext{ID: https:/orcid.org/0000-0001-7768-7320}
   \and Min Fang
      \inst{1}\footnotetext{ID: https:/orcid.org/0000-0001-8060-1321}
   \and Shaobo Zhang
      \inst{1}\footnotetext{ID: https:/orcid.org/0000-0003-2549-7247}
   \and Dengrong Lu
      \inst{1}
   \and Jixian Sun
      \inst{1}
   }

   \institute{Purple Mountain Observatory and Key Laboratory of Radio Astronomy, Chinese Academy of Sciences, \\
10 Yuanhua Road, Qixia District, Nanjing 210033, PR China; {\it lxyuan@pmo.ac.cn}\\
%% Please give the E-mail address of the author, to whom future correspondence
%% requests will be sent.  
\vs\no
   {\small Received 2025 month day; accepted 2026 month day}}

\abstract{To assess the velocity stability of CO spectral lines in the Milky Way Imaging Scroll Painting (MWISP) survey, 
we employ a cross-correlation method to measure velocity shifts across $\sim$ 10,000 CO spectra
from six reference sources observed over the ten-year duration of the survey.  
The standard deviations ($\sigma$) of these measured velocity shifts 
range from 0.03 to 0.23 km s$^{-1}$ for their $^{12}$CO lines 
and 0.03 to 0.16 km s$^{-1}$ for $^{13}$CO lines. 
We find that larger shifts are associated with broader linewidths, 
more pronounced differences between monthly and long-term variations, 
and a stronger correlation between velocity shifts of $^{12}$CO and $^{13}$CO lines. 
By examining the relation of velocity shifts with the Azimuth-Elevation of the telescope, 
as well as the velocity fields of these extended sources, 
we find that the velocity shifts exhibit systematic changes across different Azimuth-Elevation ranges. 
The patterns and amplitudes of these changes vary among sources 
and are closely linked to the extended velocity fields of sources.  
This indicates that the increased velocity shifts are primarily caused by pointing errors, 
which are more sensitive to reference sources with higher velocity gradients.  
We also provide high signal-to-noise, velocity-aligned template spectra for reference sources. 
}

   \authorrunning{Lixia Yuan, Ji Yang \& Min Fang}            %author_head in even pages
   \titlerunning{The radial velocity stability for reference sources in the MWISP survey }  % title_head in odd pages

   \maketitle
%% The author head (on even pages) and the title head (on odd pages) will be
%% automatically extracted from \author{} and \title{}. Whenever the title is too long,
%% you will be asked to supply a shorter one by inserting either \authorrunning{} or
%% \titlerunning{} before \maketitle. Anyway, you can specify your own heads.
%%
%%
%% Note: In the following text body of your manuscript, please note several differences from
%%       other major journals:
%% (1) \subsection{Please Capitalize the First Letter of Each Notional Word in Subsection Title}
%% (2) Please Capitalize the First Letter of Each Notional Word in all tables' captions

%
%________________________________________________ sections below
%
\section{Introduction}           %% first-level sections will be auto-capitalized
\label{sect:intro}
For the millimeter-wavelength observations, 
the reliable absolute calibration for the intensity of a spectral line is fundamental element \citep{Ulich1976, Kutner1981, Vanden1985, Yang1999}. 
Absolute calibration is usually conducted by the chopper-wheel method \citep{Ulich1976}, such as 
CO surveys by CfA 1.2 m telescope \citep{Dame1987, Dame2001}. 
A further check on the calibration include repeatedly observing reference spectral sources, 
whose brightness distribution must be stable in the long time. 
This ensures the line intensity scale remains consistent with previously established `standard spectra' 
and also keep a record of the system performance.
These combine instrumental calibration and empirical reference checks, 
conducted in such as Galactic Ring Surveys \citep{Jackson2006}, 
FOREST Unbiased Galactic plane Imaging survey \citep{Umemoto2017}, 
SEDIGISM survey \citep{Schuller2017}. 
That not only ensure the accuracy of the intensity scale 
but also provide a means to monitor the long-term stability of the telescope system.

The Milky Way Imaging Scroll Painting (MWISP) is an unbiased Galactic plane CO survey, 
which is conducted using the 13.7m millimeter-wavelength telescope of Purple Mountain Observatory (PMO). 
It observes $^{12}$CO, $^{13}$CO, and C$^{18}$O($J$=1-0) spectra, simultaneously \citep{Su2019, Yang2025}. 
The first phase of the MWISP CO data, which covers Galactic longitudes from l = 9.75$^{\circ}$ to 230.25$^{\circ}$ 
and Galactic latitudes from b = -5.25$^{\circ}$ to 5.25$^{\circ}$, has been publicly released \citep{Yang2025}. 
The MWISP CO survey employ a protocol that includes observations of a reference source 
immediately before and after each observation of a $30^{\prime}\times30^{\prime}$ cell. 
These reference sources have their spectral profiles monitored daily to check 
the stability of the system performance. 

In the MWISP survey, the $^{12}$CO, $^{13}$CO, and C$^{18}$O($J$=1-0) lines at 
frequencies of 115.271 GHz, 110.201 GHz, and 109.782 GHz, respectively, 
are detected simultaneously using the Superconducting Spectroscopic Array Receiver (SSAR) system \citep{Shan2012}. 
The local oscillator (LO) frequency is set to 112.6 GHz, and 
the intermediate frequency (IF) is 2.64 GHz. 
The $^{12}$CO line is placed in the upper sideband, 
while the $^{13}$CO and C$^{18}$O lines are allocated to the lower sideband. 
The first LO performs real-time Doppler tracking at the frequency of $^{12}$CO line. 
A fine-tuning frequency tracking is further conducted at the second LO to match 
the observed frequencies of $^{13}$CO and C$^{18}$O lines. 
This capability to simultaneously observe three lines with a single telescope provides 
a distinct advantage, as it helps to mitigate potential instrumental effects and 
minimizes fluctuations among the line intensities of the isotope molecule \citep{Yang1999}.

CO spectral line data can provide crucial information on kinematics and dynamics of molecular clouds, 
stellar feedback activities, and the dynamical structures of Galaxy, 
when analyzed in terms of their radial velocities with respect to the local standard of rest (LSR). 
Therefore, ensuring the reliability and accuracy of the observed radial velocities 
during an extensive CO survey is fundamental for initiating our analysis on these topics. 
Radial velocities are determined from the Doppler shift of line emission 
by comparing the measured frequency of the line emission with the rest frequency. 
This comparison allows us to establish the relative velocity along the line of sight between emission sources and receiving system. 
%This relative velocity encompasses the motion of both sources and receiving system. 
After accounting for the earth's rotation and its revolution around the Sun, 
and barycentric motion of sun system,  
we can compute radial velocities of the sources relative to the local standard of rest. 

The daily monitoring of the line profiles of the reference source during observations 
gave a direct, empirical check of the radial velocity stability. 
However, it is essential to quantitatively measure the radial velocity stability and accuracy over the extended MWISP CO survey.
In this paper, we aim to assess the stability of the observed radial velocities in the MWISP CO survey and 
identify the factors that influence the long-term stability of these velocities. 
We primarily utilize the cross-correlation method to measure the velocity shifts of 
the spectral lines from reference sources that were observed throughout the entire CO survey period. 
The structure of this paper is organized as follows:
Section 2 describes the $^{12}$CO and $^{13}$CO spectra data for reference sources.
Section 3 presents the results, including the distribution of measured velocity shifts, 
the variation of these shifts over the survey period, 
and the correlation between velocity shifts from $^{12}$CO and $^{13}$CO lines. 
Section 4 discusses the measurement uncertainties caused by the observational noise 
within the cross-correlation method. 
Also, we discuss underlying factors for the increase in velocity shifts over different reference sources.

\section{Data}
\label{sect:Obs}

The MWISP survey is an ongoing survey of carbon monoxide (CO) emissions in the northern Galactic plane, 
utilizing the 13.7m telescope situated in Delingha, China. 
A comprehensive description of the telescope's capabilities and its multibeam receiver system 
can be referenced in \cite{Su2019, Shan2012}. 
Furthermore, the observational strategy and raw data reduction processes employed in the MWISP survey 
are outlined in \cite{Su2019, Yang2025}. 
The half-power beamwidth (HPBW) of the antenna at a frequency of 115 GHz is $\sim$ 50$^{\prime \prime}$.  
The typical system temperature is roughly 250 K for $^{12}$CO lines in the upper sideband and 
around 140 K for the $^{13}$CO and C$^{18}$O lines in the lower sideband, respectively. 
This survey offers a total bandwidth of 1 GHz with 16,384 channels, 
resulting in a spectral resolution of 61 kHz per channel. 
Consequently, this configuration yields a velocity resolution of about 0.16 km s$^{-1}$ for $^{12}$CO lines 
and 0.17 km s$^{-1}$ for $^{13}$CO and C$^{18}$O lines. 
The typical root mean square (RMS) noise levels achieved for the $^{12}$CO and $^{13}$CO lines are $\sim$ 0.5 K and $\sim$ 0.3 K, respectively.

During the MWISP survey, reference sources are observed approximately every three hours 
at the beginning and end of each observational cell (0.5$^{\circ}$ $\times$ 0.5$^{\circ}$). 
The 3$\times$3 multibeam receiver uses position-switch (PS) mode to observe the reference source, 
sequentially moving from beam 1 to beam 9. 
The observation protocol for each beam follows this sequence: "Black body - OFF - ON - ON - OFF - Black body". 
Each ON and OFF observation lasts 10 seconds, while each black body observation lasts 5 seconds. 
The total time required to observe all 9 beams is about 10 minutes. 
During each observation, we obtain the spectral lines of the reference sources across beams 1 to 9 \citep{Su2019, Yang2025}.
Data preprocessing for spectral lines includes baseline fitting and the removal of bad channels. 
Any spectral lines with a poor baseline are discarded.
For the MWISP survey, we use the reference sources of L134, W51D, Sh2-140, W3(OH), NGC2264, IRC+10216, and DR21.
The basic information for these sources is listed in Table \ref{tab_bzy}. 
Among seven reference sources, source DR21 has only 184 spectral lines, 
which limited the analysis of the velocity shift variation over an extended period. 
Thus its velocity shift are not measured in this work.
In addition, for source IRC+10216, we measure the distribution of its velocity shifts, 
while this source is a late-type carbon star, whose gas emission is time-dependent and presents periodical variability \citep{Pardo2018}. 
This means velocity shifts of source IRC+10216 may be influenced by both its inherent gas motions and instrumental effects. 
Therefore, we don’t analyze its velocity shifts in relation to the velocity stability of telescope systems.

\begin{table*}
\begin{center}
\caption[]{Measured velocity shifts on the reference sources used in the MWISP survey. 
   Column 1: Name of the reference source. Columns 2 and 3:  Right Ascension (RA) and Declination (Dec) coordinates of the reference source. 
    Column 4: Standard deviation of the entire measured velocity shifts for $^{12}$CO spectral lines ($\sigma_{\rm ^{12}CO}$). 
    Column 5: The median value of the monthly standard deviations ($\sigma_{\rm m}$) of velocity shifts for $^{12}$CO spectral lines. 
    Column 6: Standard deviation of the entire measured velocity shifts for $^{13}$CO spectral lines ($\sigma_{\rm ^{13}CO}$). 
    Column 7: The median value of the monthly standard deviations of velocity shifts for $^{13}$CO spectral lines. 
    Column 8: the number of spectral lines} \label{tab_bzy}
\begin{tabular}{lccccccc}
   \hline\noalign{\smallskip}
    Source name & RA & Dec & $\sigma_{\rm ^{12}CO}$ & $\sigma_{\rm m}$ ($^{12}$CO) 
    & $\sigma_{\rm ^{13}CO}$ & $\sigma_{\rm m}$ ($^{13}$CO) & Number of Spectra \\
     & -- & -- & km s$^{-1}$ & km s$^{-1}$ &  km s$^{-1}$ & km s$^{-1}$ & -- \\
   \hline\noalign{\smallskip}
    L134 & 15h53m38.26s & -4d35m46.9s & 0.03 & 0.03 & 0.03 & 0.02 & 11205 \\
    W51D & 19h23m39.9s & 14d31m10.1s & 0.09 & 0.07 & 0.06 & 0.04 & 20807 \\
    Sh2-140 & 22h19m19.14s & 63d18m50.3s & 0.11 & 0.08 & 0.06 & 0.04 & 17120 \\
    W3(OH) & 02h27m03.88s & 61d52m24.6s & 0.23 & 0.16 & 0.16 & 0.11 & 9530 \\ 
    NGC2264 & 06h41m09.77s & 09d29m35.9s & 0.07 & 0.045 & 0.05 & 0.03 & 21752 \\
    IRC$+$10216 & 09h47m57.29 & 13d16m42.9s & 0.07 & -- & 0.12 & -- & 8685 \\
    DR21 & 20h39m01.19 & 42d19m44.9s & -- & -- & -- & -- & 184 \\
    \noalign{\smallskip}\hline
\end{tabular}
\end{center}
\end{table*}

\section{Results}
\subsection{Method}
The cross-correlation function (CCF) measures the similarity between two spectra 
and is particularly sensitive to velocity shifts when two spectra are identical.  
This method is commonly used to measure velocity shifts in galaxies \citep{Tonry1979, York2000} and stars \citep{Katz2004}. 
In this study, we apply the cross-correlation function to determine the velocity shifts among 
the spectral lines of each reference source observed during the 10-year period of the MWISP phase I survey. 
Our observed spectra are sampled with a spectral resolution of $\sim$ 0.16 km s$^{-1}$ per channel. 
To analyze the data, we employ the discrete form of CCF, given by the equation: 
\begin{equation}
    CCF(\Delta x)=\Sigma_{x=1}^{N}S(x)T(x+\Delta x)dx.
\end{equation}
where S(x) represents the spectrum as a function of velocity channels, 
while T(x) denotes the template spectrum. 
The term $\Delta$x refers to the lag in the CCF, 
which reaches its maximum value when $\Delta$x aligns with both functions.
To enhance measurement accuracy to sub-pixel levels (approximately 1/10 of a pixel), 
we apply a second-order polynomial to model the peak of the cross-correlation function 
within a defined window of ($\Delta$x-1, $\Delta$x+1). 

\subsection{Templates of reference sources}
To obtain accurate velocity shift measurements using the CCF method, 
it is essential to have a low-noise template. 
One effective way to improve the signal-to-noise (S/N) ratio of a spectral line profile 
is to utilize the average shape of a typical spectral line. 
We begin by selecting one spectrum as the initial template to measure the velocity shifts 
between this spectrum and others. 
Next, we align all spectra with this template based on the measured velocity shifts.  
Once aligned, we average all the spectral lines to create what we define as the second template. 
Using this second template, we then recalculate the velocity shifts between it and other spectra. 
After using the second template to measure the velocity shifts,
we then create a third template. 
This third template is another mean spectrum, 
aligned using the velocity shifts derived from the second template. 
It is important to note that the measured velocity shifts are relative values, 
representing how much each line profile shifts compared to the template. 
The first template, being just one of the observed spectra, 
may not accurately represent the true velocity of source. 
In statistical terms, the median values of measured velocity shifts represent 
the shift of the template in relation to the source's accurate velocity. 
To refine this, we adjust the third template by applying the median value of the measured 
velocity shifts obtained from the second template. 
This adjustment results in a final template, 
whose line profile aligns most closely with the true velocity of the source.

\begin{figure*}[ht]
   \centering
   \includegraphics[width=15cm, angle=0]{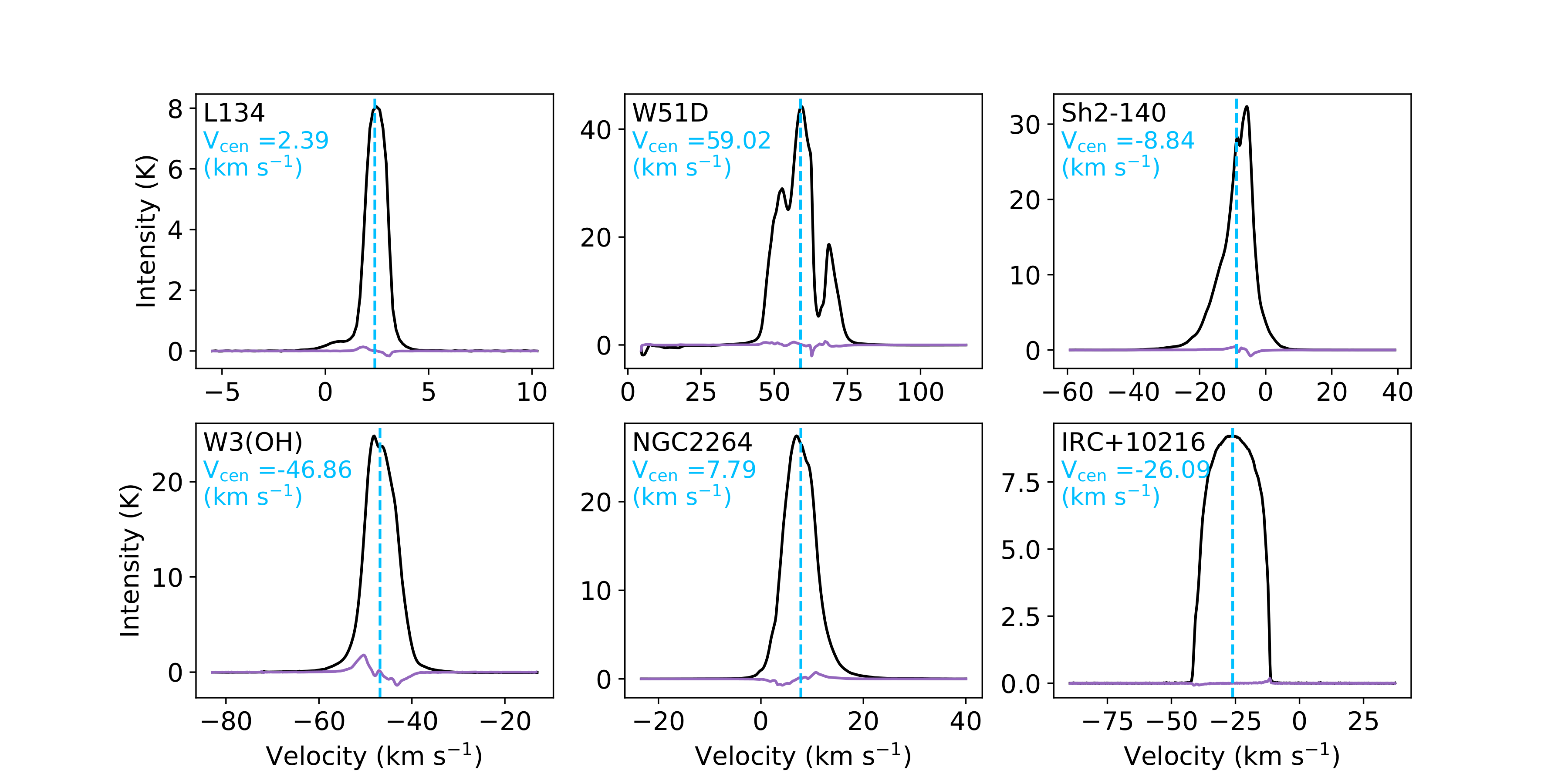}
   \caption{The aligned-mean $^{12}$CO spectral line profile for each reference source,  
   with the centroid velocity marked by a blue-dashed line and noted in the upper left corner of each plane. 
   The purple lines represent the residuals between the aligned-mean spectrum and the mean spectrum.}
   \label{fig:f_12CO}
\end{figure*}

\begin{figure*}[ht]
   \centering
   \includegraphics[width=15cm, angle=0]{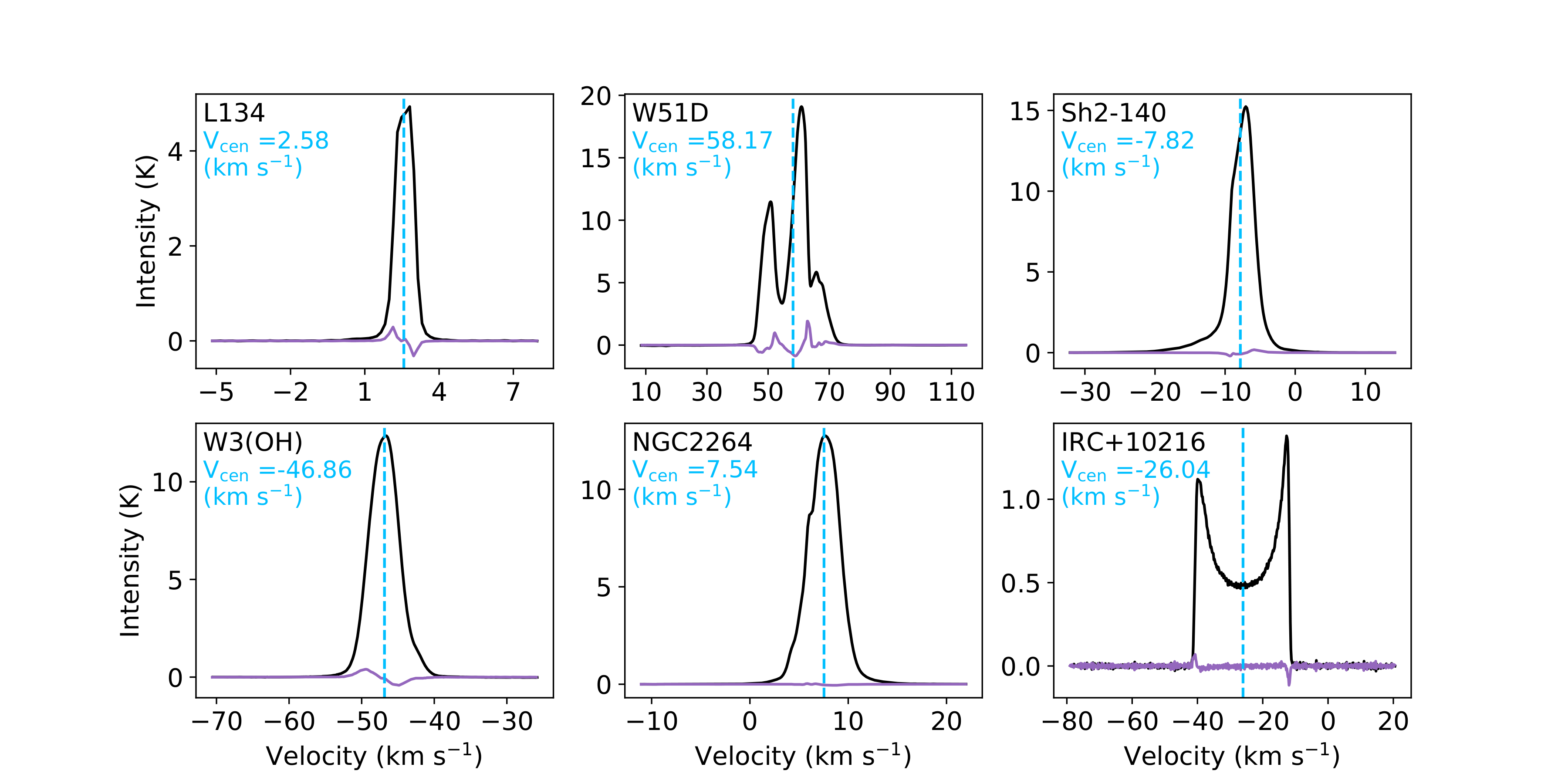}
   \caption{Same as Figure \ref{fig:f_12CO}, but for the $^{13}$CO spectral lines of reference sources.}
   \label{fig:f_13CO}
\end{figure*}

Figure \ref{fig:f_12CO} and \ref{fig:f_13CO} illustrate the final templates for 
the $^{12}$CO and $^{13}$CO spectral lines corresponding to each reference source. 
Additionally, we also present the residuals between the aligned-mean spectrum and the mean spectrum, 
which represents the average profile of the entire spectra without taking velocity shifts into account. 
There are observable differences between the two mean spectra. 
The aligned-mean spectra integrate over ten thousand lines observed over a period of 10 years, 
significantly reducing the impact of noise and velocity shifts.  
This process yields a more statistically accurate line profile for these reference sources. 
The centroid velocities for each source, 
calculated from the aligned-mean $^{12}$CO and $^{13}$CO spectral lines,
are also indicated in Figures \ref{fig:f_12CO} and \ref{fig:f_13CO}. 

\subsection{Distribution of measured velocity shifts}
\begin{figure*}[ht]
   \centering
   \includegraphics[width=15cm, angle=0]{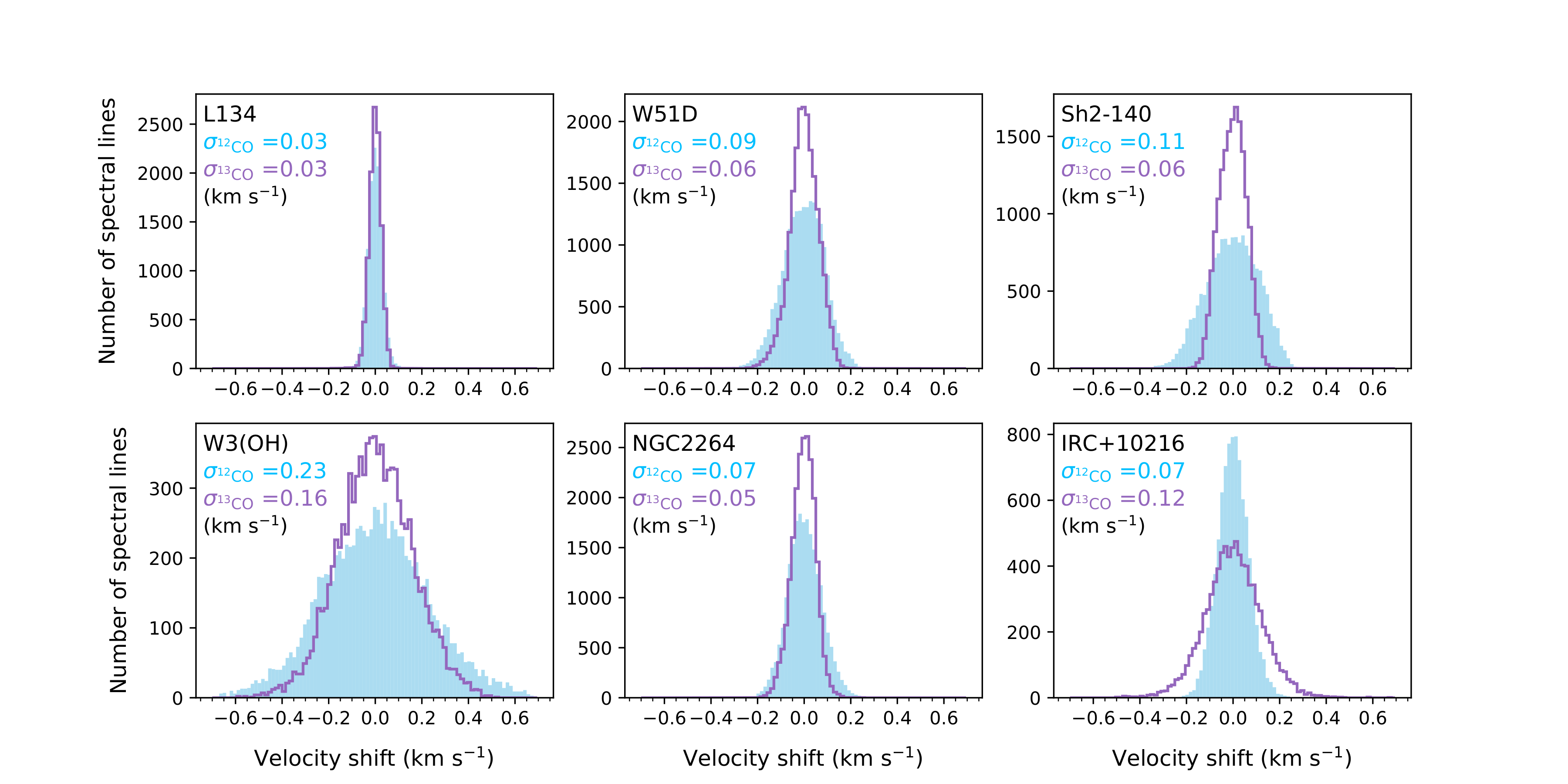}
   \caption{The distribution of measured velocity shifts among $^{12}$CO spectral lines (blue histogram) 
   and $^{13}$CO spectral lines (purple histogram) for each reference source. 
   These velocity shifts have been adjusted by subtracting their median values. 
   The corresponding standard deviations of velocity shifts ($\sigma_{\rm ^{12}CO}$ for $^{12}$CO lines 
   and $\sigma_{\rm ^{13}CO}$ for $^{13}$CO lines) are indicated in each plane.}
   \label{fig:vshift}
\end{figure*}

Figure \ref{fig:vshift} illustrates the distributions of measured velocity shifts of 
$^{12}$CO and $^{13}$CO spectral lines for each reference source, 
along with their corresponding standard deviation ($\sigma$) values. 
We should note that the measured velocity shifts are relative values compared to templates.
In statistical terms, the median values of measured velocity shifts represent 
the shift of the template in relation to the source's accurate velocity.
Thus the velocity shifts values have been adjusted by subtracting their median values.  
For $^{12}$CO spectral lines, the $\sigma_{\rm ^{12}CO}$ values vary by sources, 
ranging from 0.03 to 0.23 km s$^{-1}$.
The source identified as L134 has the lowest $\sigma_{\rm ^{12}CO}$ value at 0.03 km s$^{-1}$. 
For source NGC2264, W51D and Sh2-140, 
the $\sigma_{\rm ^{12}CO}$ value is 0.07, 0.09, and 0.11 km s$^{-1}$, respectively. 
The source W3(OH) shows the highest $\sigma_{\rm ^{12}CO}$ value at 0.23 km s$^{-1}$. 
For $^{13}$CO spectral lines, 
the $\sigma_{\rm ^{13}CO}$ values for the velocity shifts range from 0.03 to 0.16 km s$^{-1}$. 
The $\sigma_{\rm ^{13}CO}$ for L134 is 0.026 km s$^{-1}$, which is close to the $\sigma_{\rm ^{12}CO}$ value. 
In comparison, the $\sigma_{\rm ^{13}CO}$ value is 0.05 km s$^{-1}$ for NGC2264 and 
0.06 km s$^{-1}$ for both sources W51D and Sh2-140. 
Their values are lower about 0.02 -- 0.05 km s$^{-1}$ than their corresponding $\sigma_{\rm ^{12}CO}$ values. 
For W3(OH) source, its $\sigma_{\rm ^{13}CO}$ value is 0.16 km s$^{-1}$, 
obviously lower about 0.07 km s$^{-1}$ than its $\sigma_{\rm ^{12}CO}$. 
These $\sigma_{\rm ^{13}CO}$ values are all lower than their corresponding $\sigma_{\rm ^{12}CO}$ values. 
Additionally, the line widths of $^{13}$CO spectral lines are narrower than 
those of $^{12}$CO lines. 
This suggests that the velocity accuracy for the sources may be related to the widths of their spectral lines. 

\begin{figure*}[ht]
   \centering
   \includegraphics[width=15cm, angle=0]{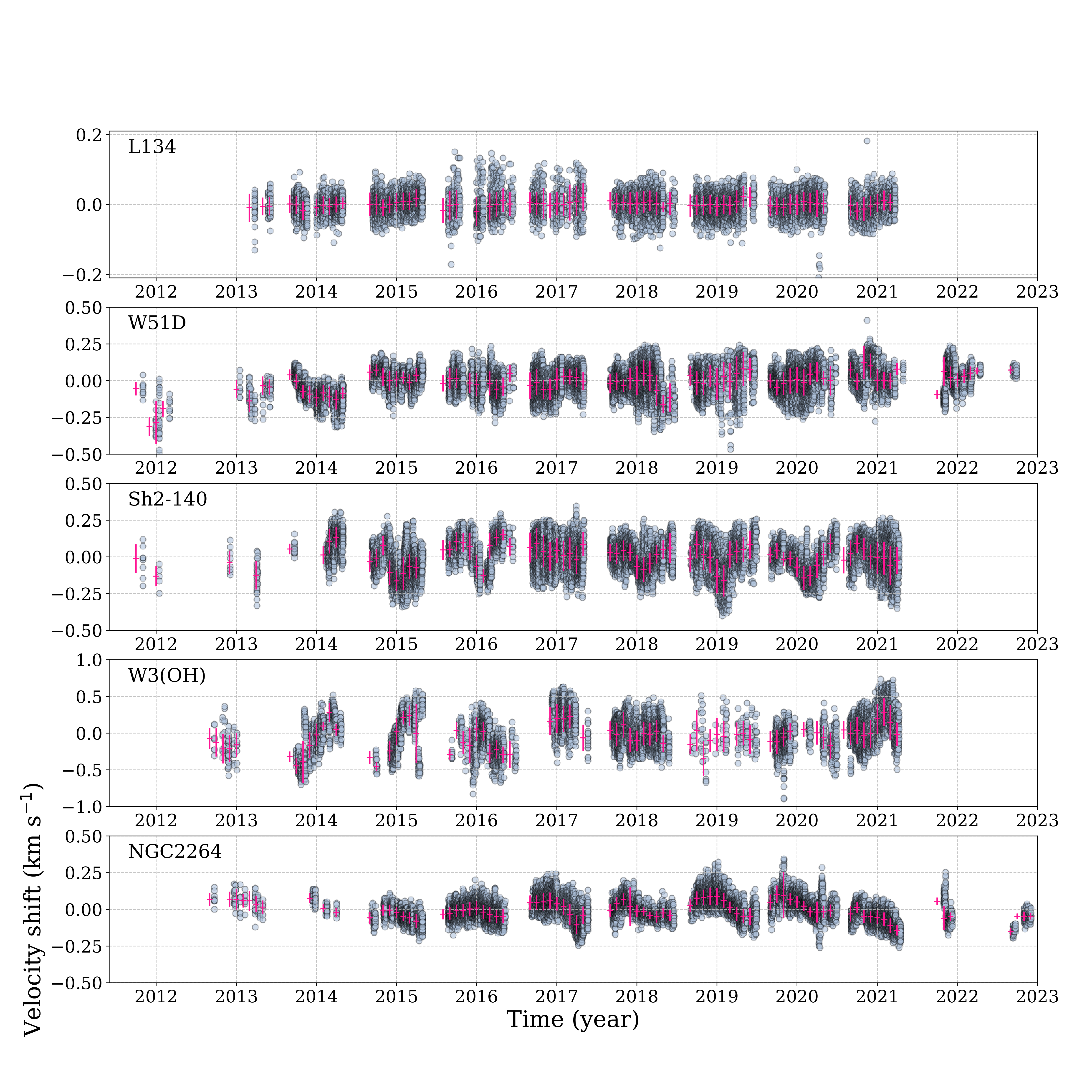}
   \caption{The variation of velocity shifts among $^{12}$CO spectral lines for each reference source 
   across the MWISP survey period. The pink bars represent the standard deviations of the 
   monthly velocity shifts and the crosses indicate the median values of these monthly shifts.}
   \label{fig:time_12CO}
\end{figure*}

\begin{figure*}[ht]
   \centering
   \includegraphics[width=15cm, angle=0]{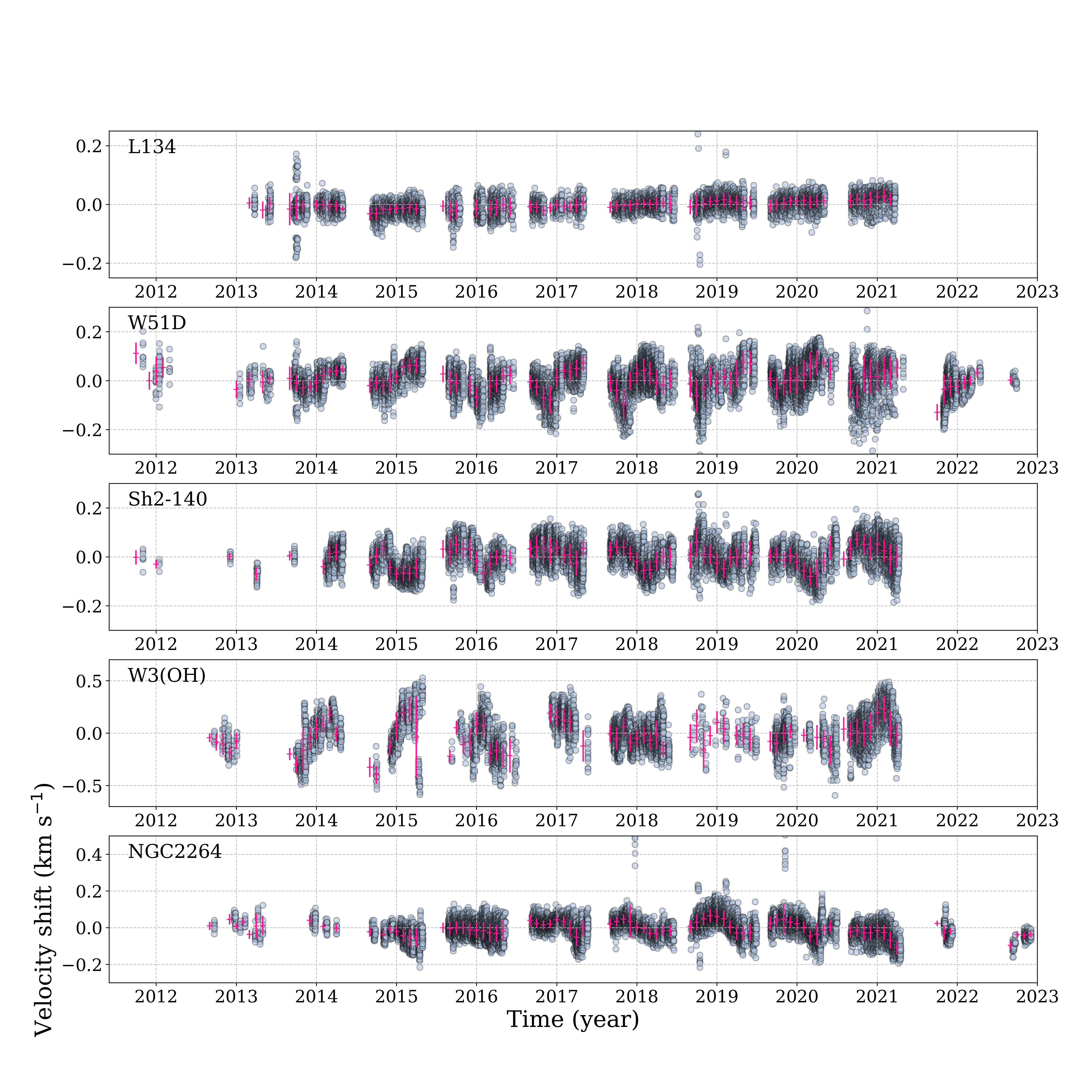}
   \caption{The variation of velocity shifts among $^{13}$CO spectral lines for each reference source 
   across the MWISP survey period. The pink bars represent the standard deviations of the 
   monthly velocity shifts and the crosses indicate the median values of these monthly shifts.}
   \label{fig:time_13CO}
\end{figure*}

The reference sources observed during the MWISP survey period spans from 2012 to 2023. 
We also illustrate the variation of velocity shifts for each reference source across the survey period. 
Figures \ref{fig:time_12CO} and \ref{fig:time_13CO} display the changes in velocity shifts 
for the $^{12}$CO and $^{13}$CO spectral lines, respectively. 
We find that these velocity shift variations exhibit periodic fluctuations 
that occur over the span of one year. 
For reference source L134, the majority of velocity shifts range between $\pm$0.1 km s$^{-1}$ for 
both the $^{12}$CO and $^{13}$CO lines. 
For reference sources W51D, Sh2-140, and NGC2264, their velocity shifts primarily fluctuate 
within $\pm$0.3 km s$^{-1}$ for $^{12}$CO lines and within $\pm$0.2 km s$^{-1}$ for $^{13}$CO lines. 
For the W3(OH) source, the velocity shifts range between $\pm$0.7 km s$^{-1}$ for $^{12}$CO lines 
and $\pm$0.5 km s$^{-1}$ for $^{13}$CO lines. 

To assess whether the yearly variations are influenced by the Earth's motion, 
we compare the Doppler corrections derived from three independent algorithm. 
One is the \href{http://www.starlink.ac.uk/docs/sun67.htx/sun67.html}{SLALIB}, 
used in the MWISP CO survey, the other two is the \href{http://astro.uni-tuebingen.de/software/idl/astrolib/astro/baryvel.pro}{baryvel} procedure \citep{Stumpff1980} from IDL 
and the radial\_velocity\_correction procedure of 
\href{https://docs.astropy.org/en/stable/api/astropy.coordinates.CIRS.html}{astropy.coordinates.CIRS}. 
The differences in barycentric corrections among these three procedures are typically less than 0.01 km s$^{-1}$. 
As illustrated in Figure \ref{fig:f_vbary}, we present barycentric corrections from the SLALIB and 
baryvel.pro procedures for sources W3(OH) observed between October 2014 and May 2015. 
The differences between two methods are about 0.008 km s$^{-1}$.
During this period, the measured velocity shifts ranged from -0.5 to 0.5 km s$^{-1}$. 
Additionally, the velocity shift $\sigma$ values varied from 0.03 to 0.23 km s$^{-1}$ for different sources, 
indicating that the Doppler correction was not the primary factor influencing these shifts.

%These fluctuation patterns of $^{12}$CO spectral lines closely resemble those of the $^{13}$CO spectral lines. 
%The reasons behind these fluctuations will be further discussed in Section \ref{sec:reason}.

We also present the monthly standard deviation of velocity shifts, 
represented by pink bars in Figure \ref{fig:time_12CO} and \ref{fig:time_13CO}. 
This monthly measure reflects temporal variations over a localized period of one month, 
minimizing the influence of fluctuations that occur over a year. 
The median values for these monthly standard deviations, denoted as $\sigma_{\rm m}$, are listed in Table \ref{tab_bzy}.
Furthermore, we compare $\sigma_{\rm m}$ with the standard deviation of velocity shifts for the 
entire survey period, referred to as $\sigma$. 

\begin{figure*}[ht]
   \centering
   \includegraphics[width=15cm, angle=0]{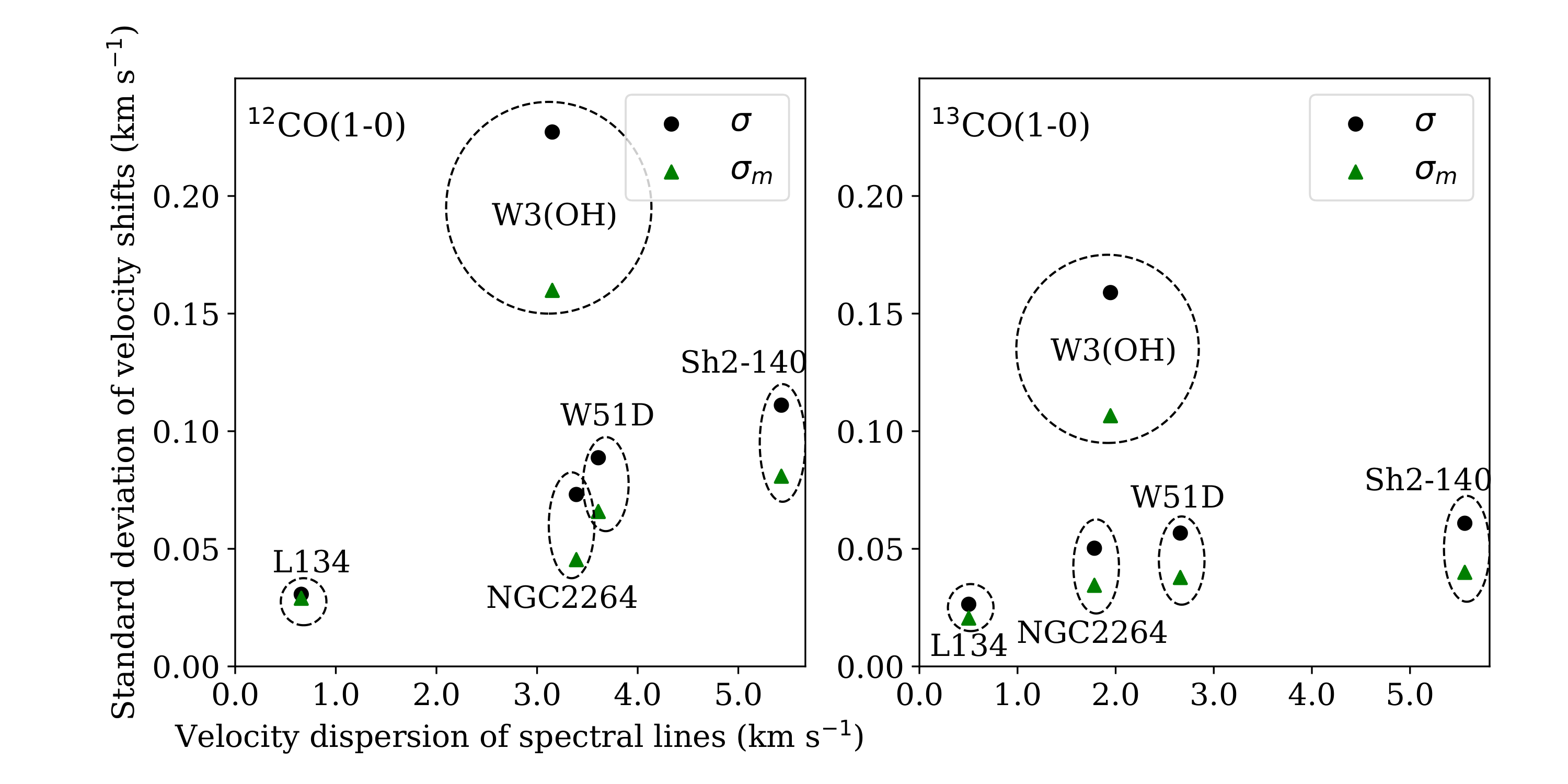}
   \caption{Relation between standard deviation of velocity shifts and velocity dispersion of $^{12}$CO (left panel) and $^{13}$CO (right panel) spectral lines. 
   The $\sigma$ represents the standard deviation of velocity shifts over the entire survey period. 
   The $\sigma_{\rm m}$ denotes the median value for the monthly standard deviations of velocity shifts.}
   \label{fig:vshift_vsigma}
\end{figure*}

Figure \ref{fig:vshift_vsigma} illustrates the values of $\sigma$ and $\sigma_{\rm m}$ 
in relation to the velocity dispersions of spectral lines for reference sources. 
Firstly, we observe a trend where velocity shifts increase alongside the velocity dispersions of spectral lines. 
However, the source W3(OH) stands out, as its velocity shifts are significantly larger for 
both the $^{12}$CO(1-0) and $^{13}$CO(1-0) lines.  
This suggests that additional factors may be contributing to its larger shifts. 
Moreover, the differences between $\sigma$ and $\sigma_{\rm m}$ vary across different sources.  
For source L134, the values of $\sigma$ and $\sigma_{\rm m}$ are very close. 
In contrast, for the sources NGC2264, W51D, and Sh2-140, 
the differences between $\sigma$ and $\sigma_{\rm m}$ are about 0.02 -- 0.03 km s$^{-1}$. 
For source W3(OH), the difference is 0.07 km s$^{-1}$ for $^{12}$CO lines and 0.05 km s$^{-1}$ for $^{13}$CO lines.   
We find that some sources have greater $\sigma$ values, 
and the disparity between $\sigma$ and $\sigma_{\rm m}$ tends to be larger in these cases.
%While, the differences of $\sigma$ and $\sigma_{m}$ take a percentage of 20-30$\%$ of $\sigma$.
This indicates that the factors contributing to velocity shifts can be categorized into two models: 
one affecting in a localized period of less than one month and the other spanning one year. 
For source W3(OH), the one-year fluctuations significantly impact on the velocity shifts. 
In the cases of sources NGC2264, W51D, and Sh2-140, 
there is some influence from factors over one year, it is not as pronounced. 
For source L134, the nearly identical values of $\sigma$ and $\sigma_{\rm m}$ 
imply that the localized-time factors are the primary contributors to the observed shifts.

\subsection{The correlation between velocity shifts of $^{12}$CO and $^{13}$CO spectral lines}

The $^{12}$CO lines are positioned in the upper sideband, 
while the $^{13}$CO lines are allocated to the lower sideband.
This requires a fine-tuning at the second LO to ensure frequency tracking.   
By comparing the velocity shift variations of the $^{12}$CO and $^{13}$CO lines, 
we can determine whether the factors causing these shifts stem from systematic factors related to telescope's observation 
or accuracy focusing on the Doppler tracking.

\begin{figure*}[ht]
   \centering
   \includegraphics[width=15cm, angle=0]{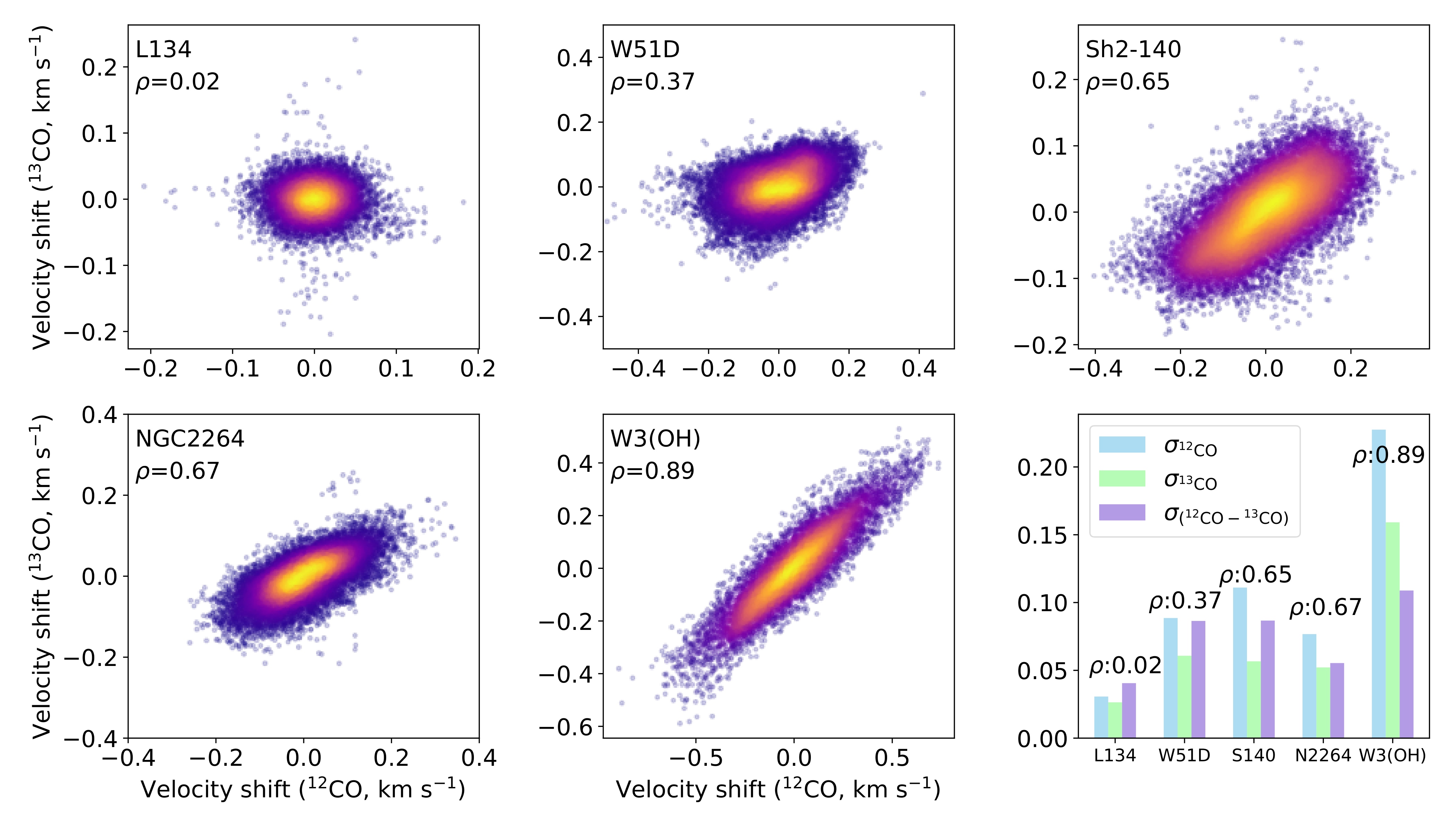}
   \caption{Correlation between $^{12}$CO velocity shifts and $^{13}$CO velocity shifts for each source. 
   The corresponding Spearman correlation coefficient ($\rho$) is indicated in each panel. 
   In the lower-right panel, the bars represent the standard deviations of velocity shifts for 
   $^{12}$CO lines ($\sigma_{\rm ^{12}CO}$) and $^{13}$CO lines ($\sigma_{\rm ^{12}CO}$), 
   as well as the standard deviation for the differences between velocity shifts of $^{12}$CO and 
   $^{13}$CO lines ($\sigma_{\rm (^{12}CO-^{13}CO)}$), 
   the label `S140' refer to source Sh2-140 and the `N2264' refer to source NGC2264.}
   \label{fig:vshift_corr}
\end{figure*}

Figure \ref{fig:vshift_corr} illustrates the correlations between velocity shifts of $^{12}$CO ($\Delta V_{\rm ^{12}CO}$) and 
$^{13}$CO ($\Delta V_{\rm ^{13}CO}$) for each reference source, along with their Spearman correlation coefficients ($\rho$). 
We find that correlations between $\Delta V_{\rm ^{12}CO}$ and $\Delta V_{\rm ^{13}CO}$ differ across various sources. 
Additionally, we estimate the standard deviation of the differences between $\Delta V_{\rm ^{12}CO}$ and $\Delta V_{\rm ^{13}CO}$, 
denoted as $\sigma_{\rm (^{12}CO-^{13}CO)}$ = $\sigma(\Delta V_{\rm ^{12}CO} - \Delta V_{\rm ^{13}CO}$). 
The comparison of $\sigma_{\rm ^{12}CO}$, $\sigma_{\rm ^{13}CO}$, and $\sigma_{\rm (^{12}CO-^{13}CO)}$ for each source 
is also presented in Figure \ref{fig:vshift_corr}. 
There appears to be a trend indicating that the correlation between $\Delta V_{\rm ^{12}CO}$ and $\Delta V_{\rm ^{13}CO}$ is stronger, 
with the values of $\sigma_{\rm ^{12}CO}$ or $\sigma_{\rm ^{13}CO}$ typically being greater than $\sigma_{\rm (^{12}CO-^{13}CO)}$ values. 
For sources L134 and W51D, the $\Delta V_{\rm ^{12}CO}$ and $\Delta V_{\rm ^{13}CO}$ show weak correlations, 
with $\rho$ values of 0.02 and 0.37, respectively. 
The $\sigma_{\rm (^{12}CO-^{13}CO)}$ values for these sources are comparable to their $\sigma_{\rm ^{12}CO}$ values.
In contrast, Sh2-140 and NGC2264 exhibit moderate correlations, with $\rho$ values of 0.65 and 0.67, respectively. 
Their $\sigma_{\rm (^{12}CO-^{13}CO)}$ values are lower than their $\sigma_{\rm ^{12}CO}$ values but higher than their $\sigma_{\rm ^{13}CO}$ values.
The source W3(OH) has a strong correlation, with a $\rho$ value of 0.89,
and its $\sigma_{\rm (^{12}CO-^{13}CO)}$ values are lower than both $\sigma_{\rm ^{12}CO}$ and $\sigma_{\rm ^{13}CO}$ values. 
This suggests that the higher velocity shifts of W3(OH) are significantly influenced by the systematic factors related to the telescope observation, 
which induce both $^{12}$CO and $^{13}$CO velocity shifts simultaneously.
For Sh2-140 and NGC2264, the velocity shifts are somewhat influenced by these telescope systematic factors,
while for L134, its velocity shift of $^{12}$CO and $^{13}$CO lines primarily reflect the 
Doppler tracking accuracy.

\section{Discussion}
\subsection{Measurement uncertainty}
These observed spectral lines exhibit noise and display different line profiles 
depending on the reference sources, with even slight changes in intensity. 
These factors can affect the accuracy of velocity shift measurements.  
To ensure the measurement uncertainty on velocity shifts,  
we utilize the template for the $^{12}$CO line of L134 to assess 
how these factors influence measurement accuracy. 
First, we generate 1000 spectral lines that have uniform line profiles with the template, 
but introduce random velocity shifts that follow a normal distribution. 
These shifts have a mean ($\mu$) of 0.0 km s$^{-1}$ and a standard deviation ($\sigma$) of 0.05 km s$^{-1}$. 
Next, we apply the CCF to measure the velocity shifts between these simulated lines and the template. 
We find that the measured velocity shifts have a mean value of 0.003 km s$^{-1}$ and $\sigma$ value of 0.05 km s$^{-1}$. 
This result closely aligns with the simulated distribution, 
indicating a measurement accuracy of 0.003 km s$^{-1}$.

\begin{figure*}[ht]
   \centering
   \includegraphics[width=15cm, angle=0]{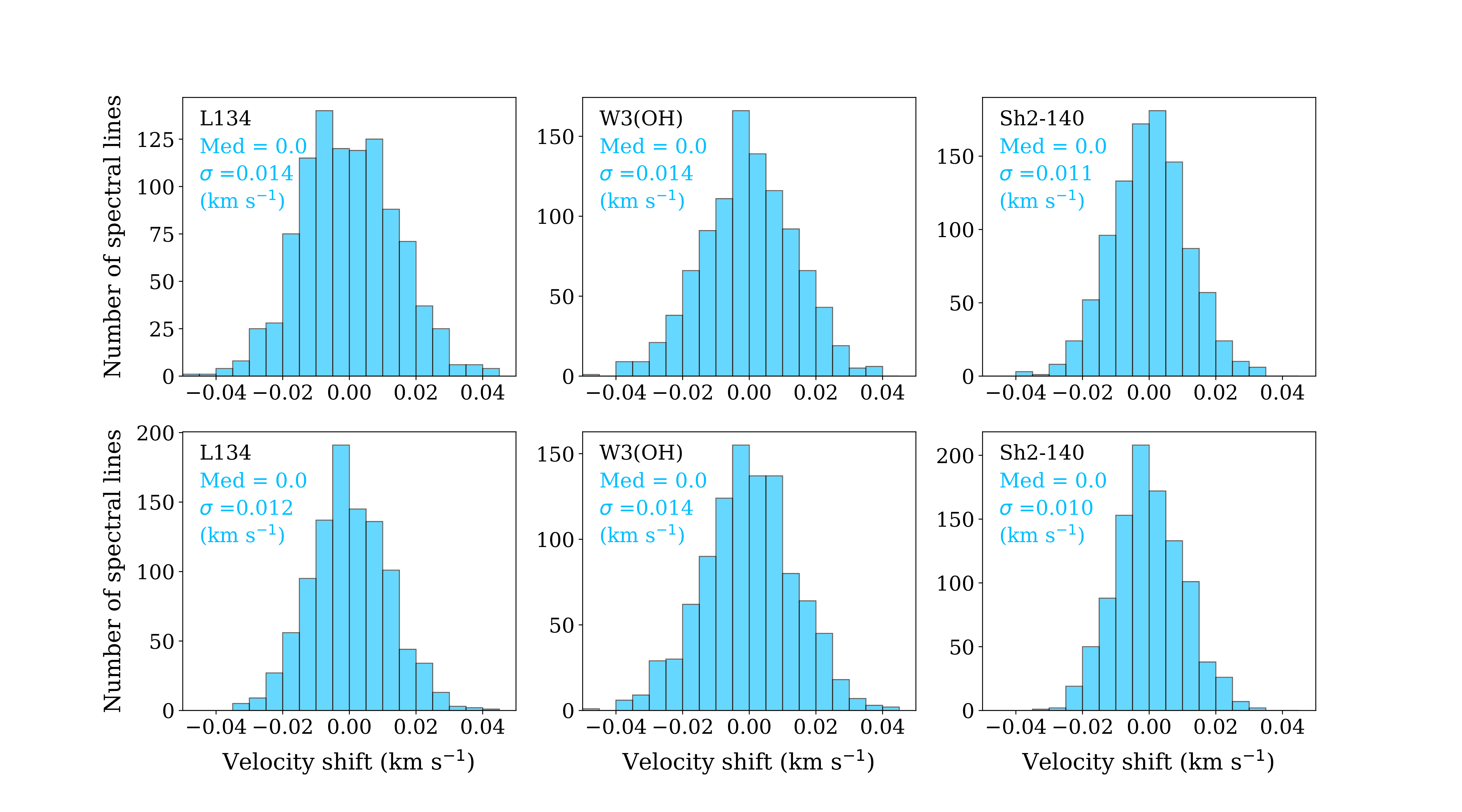}
   \caption{Distribution of measured velocity shifts caused by the observational noise on $^{12}$CO (upper plane) and $^{13}$CO (lower plane) 
   spectral profiles for source L134, W3(OH), and Sh2-140.
   The corresponding median (Med) and standard deviation ($\sigma$) values of these velocity shifts are indicated in each panel.}
   \label{fig:f_test}
\end{figure*}

The observed spectral lines contain observational noises.  
To investigate how these noises affect the resultant velocity shifts of the CCF, 
we generate 1,000 spectral lines identical to the template for the $^{12}$CO line of L134, 
meanwhile introduce noise characterized by a normal distribution with $\mu$ of 0.0 and $\sigma$ of 0.5, 
which reflects the typical noise for the $^{12}$CO spectral lines from the MWISP survey.
After calculating the velocity shifts between these simulated lines and the template,
Figure \ref{fig:f_test} shows the distribution of the measured velocity shifts and 
the $\sigma$ value is 0.014 km s$^{-1}$, which is measurement uncertainty induced by the noise.
Additionally, we test other reference sources that exhibit different line profiles 
and widths. Using the templates from these sources, we generated another 1000 spectral lines 
that replicate their templates. 
We again introduced noise characterized by a normal distribution with $\mu$ of 0.0 and $\sigma$ of 0.5 for $^{12}$CO(1-0) templates, 
as well as a normal distribution of $\mu$=0.0 and $\sigma$=0.25 for their $^{13}$CO(1-0) templates. 
After measuring the velocity shifts between the templates of various sources and the corresponding simulated samples, 
we found that their standard deviations ranged from 0.007 to 0.014 km s$^{-1}$ across different reference sources. 
Specifically, the source W51D has the lowest value of 0.007 km s$^{-1}$, 
while source L134 and W3(OH) exhibit the highest value of 0.014 km s$^{-1}$.
Figure \ref{fig:f_test} illustrates the distribution of velocity shifts 
caused by the noise for source L134, W3(OH), and Sh2-140.  

We further analyze how changes in intensity influence the measurement of velocity shifts. 
The standard deviation of the integrated intensity changes of reference sources is about 5$\%$.
Aside from the added noise, the intensities of line samples change at the scale 
with a normal distribution of $\mu$=0.0 and $\sigma$=0.05. 
As a result, the velocity shifts exhibit a uniform standard deviation with 
that calculated from line samples that introduce only noise. 
This suggests the overall change in line intensities do not significantly impact the measured velocity shifts. 
However, during actual observations, the line intensity may not change uniformly across the entire line profile, 
which is typically related to the velocity. 

After accounting for noise, variations in line intensities, and different line profiles, 
we find that the accuracy of the measured velocity shifts for our reference sources 
is about 0.04 km s$^{-1}$. 
This value is approximately three times the standard deviation of velocity shifts caused by the noise. 

\subsection{Underlying factors for the velocity shifts \label{sec:reason}}

The velocity shift modes across reference sources are different. 
First, source L134, which has the lowest $\sigma$ value of 0.03 km s$^{-1}$, 
its $\sigma_{\rm m}$ and $\sigma$ are nearly consistent. 
Additionally, the velocity shifts for $^{12}$CO and $^{13}$CO lines show a weak correlation. 
For source Sh2-140, NGC2264 and W51D, 
their $\sigma_{\rm ^{12}CO}$ range from 0.07 to 0.11 km s$^{-1}$ and $\sigma_{\rm ^{13}CO}$ are about 0.06 km s$^{-1}$. 
Furthermore, the $\sigma$ values for both $^{12}$CO and $^{13}$CO lines 
are higher by 0.02 km s$^{-1}$ than their respective $\sigma_{\rm m}$ values. 
Moreover, the velocity shifts for $^{12}$CO and $^{13}$CO lines show a moderate correlation in these sources. 
Lastly, for source W3(OH), which has the highest $\sigma$ value of 0.23 km s$^{-1}$ for $^{12}$CO lines and 
0.16 km s$^{-1}$ for $^{13}$CO lines. 
The difference between $\sigma$ and $\sigma_{\rm m}$ is also significant, 
with a value of 0.07 km s$^{-1}$ for $^{12}$CO lines and 0.05 km s$^{-1}$ for $^{13}$CO lines.
Its correlation between velocity shifts for $^{12}$CO and $^{13}$CO lines is strong. 
Overall, as the velocity shifts of sources increase, 
the spectral lines of these sources tend to have wider linewidths. 
Additionally, the differences between $\sigma$ and $\sigma_{\rm m}$ tend to become more pronounced,
and the correlations between velocity shifts for $^{12}$CO lines and those for $^{13}$CO lines tend to be stronger.
These suggest that, from source L134 to W3(OH), 
systematic factors related to telescope's observation induce the simultaneous increase in 
the velocity shifts of both $^{12}$CO and $^{13}$CO lines.

In actual observations, several factors,  
such as ambient temperature and the Azimuth and Elevation coordinates of the telescope, 
can influence the accuracy of telescope observations. 
To understand the underlying factors that induced the velocity shifts, 
we collect the observational elements, 
including the Azimuth (Az) and Elevation (El) coordinates of reference sources, 
ambient temperature (T$_{\rm amb}$), and system temperature (T$_{\rm sys}$) for the observed spectral lines. 
Utilizing a 10-year dataset of these observational variables from the MWISP survey offers 
significant advantages in quantifying their impact 
on the observed velocity shifts over the long term. 
This information could be crucial for enhancing telescope facilities and 
improving their observational accuracy.

\subsubsection{The distribution of centroid velocities on the extended reference sources}
Since the velocity shift modes are related to reference source themselves, 
so we briefly introduce the basic characters on these sources. 
First, these reference sources are extended sources, 
which are specific positions having strong CO emission within molecular clouds. 
The source L134 is located within the dark cloud Lynds 134, 
an isolated high Galactic latitude dark cloud with a distance of about 150 pc \citep{Mahoney1976}. 
The W51D source is sited within W51 giant molecular cloud, which lies at a distance of 5.41$^{+0.31}_{-0.28}$ kpc \citep{Sato2010}. 
This region is an active star-forming complex that contains multiple HII regions and embedded clusters \citep{Carpenter1998, Parsons2012, Ginsburg2017}. 
The NGC2264 source is within an active cluster-forming region populated by hundreds of young stellar objects 
\citep{Dahm2008, Wang2024}, situated in the Mon OB1 molecular cloud complex. 
It has a distance of 715$^{+81}_{-42}$ pc, as determined by Gaia DR2 \citep{Zucker2020}. 
The W3(OH) source is associated with an ultracompact (UC) HII region \citep{Reid1995} 
and is located in the W3 giant molecular complex at a distance of 2 kpc \citep{Lada1978, Yamada2024}.

\begin{figure*}[ht]
   \centering
   \includegraphics[width=14cm, angle=0]{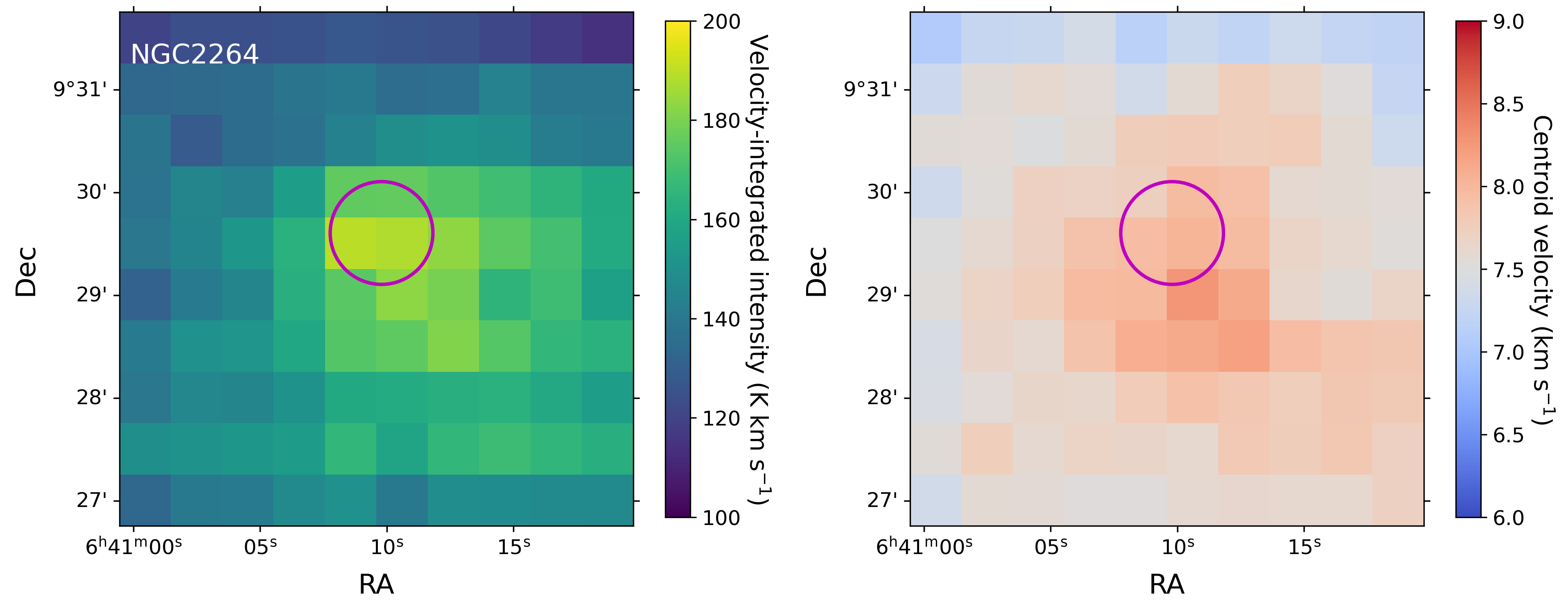}
   \includegraphics[width=14cm, angle=0]{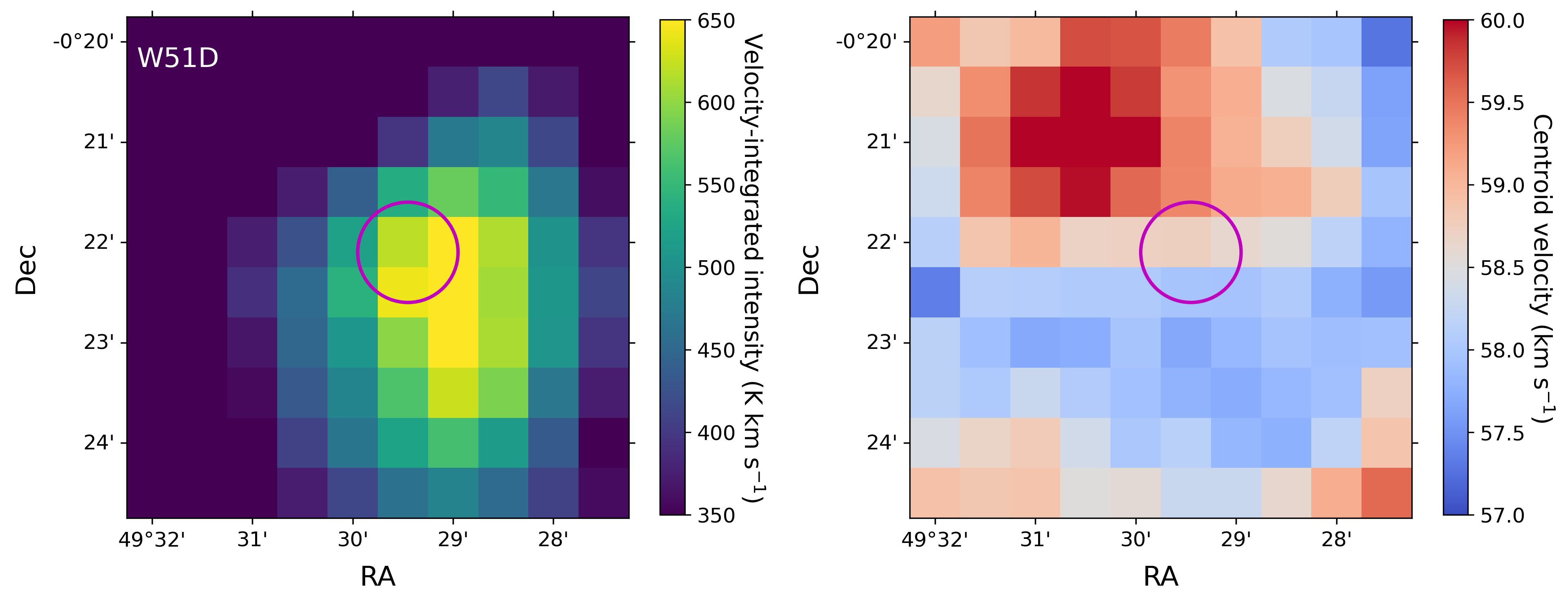}
   \includegraphics[width=14cm, angle=0]{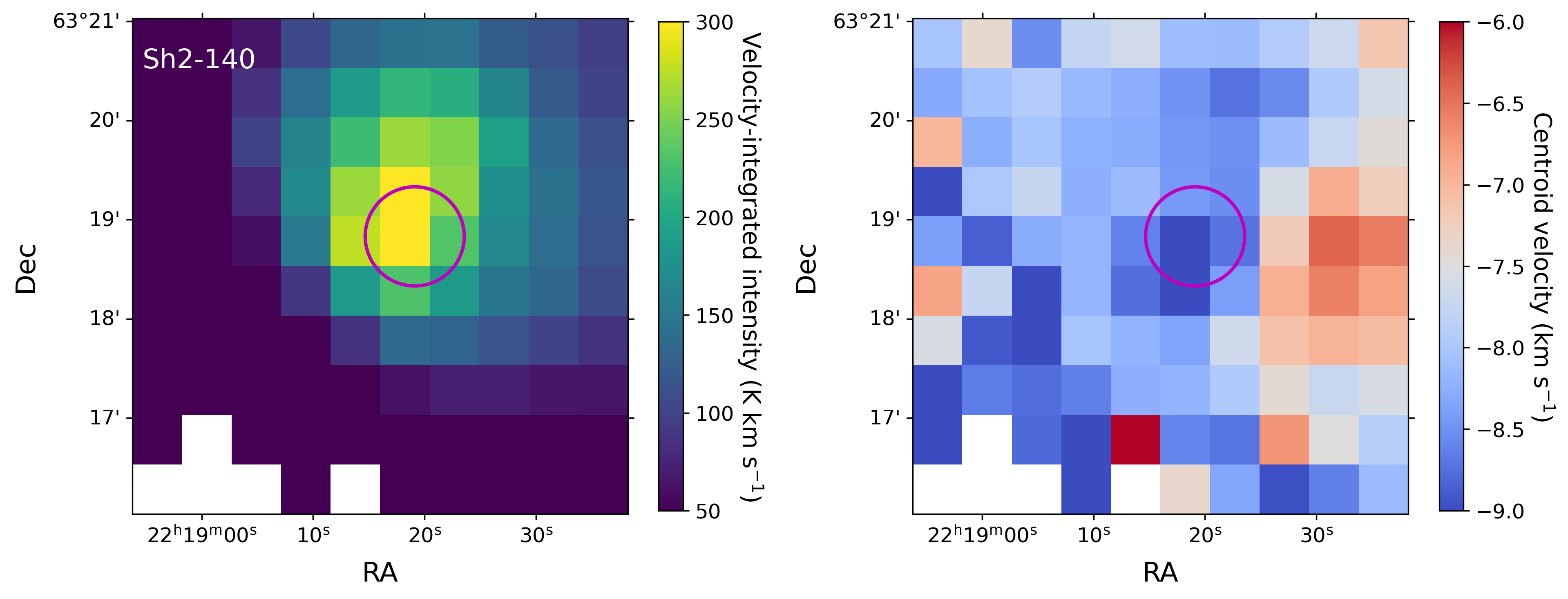}
   \includegraphics[width=14cm, angle=0]{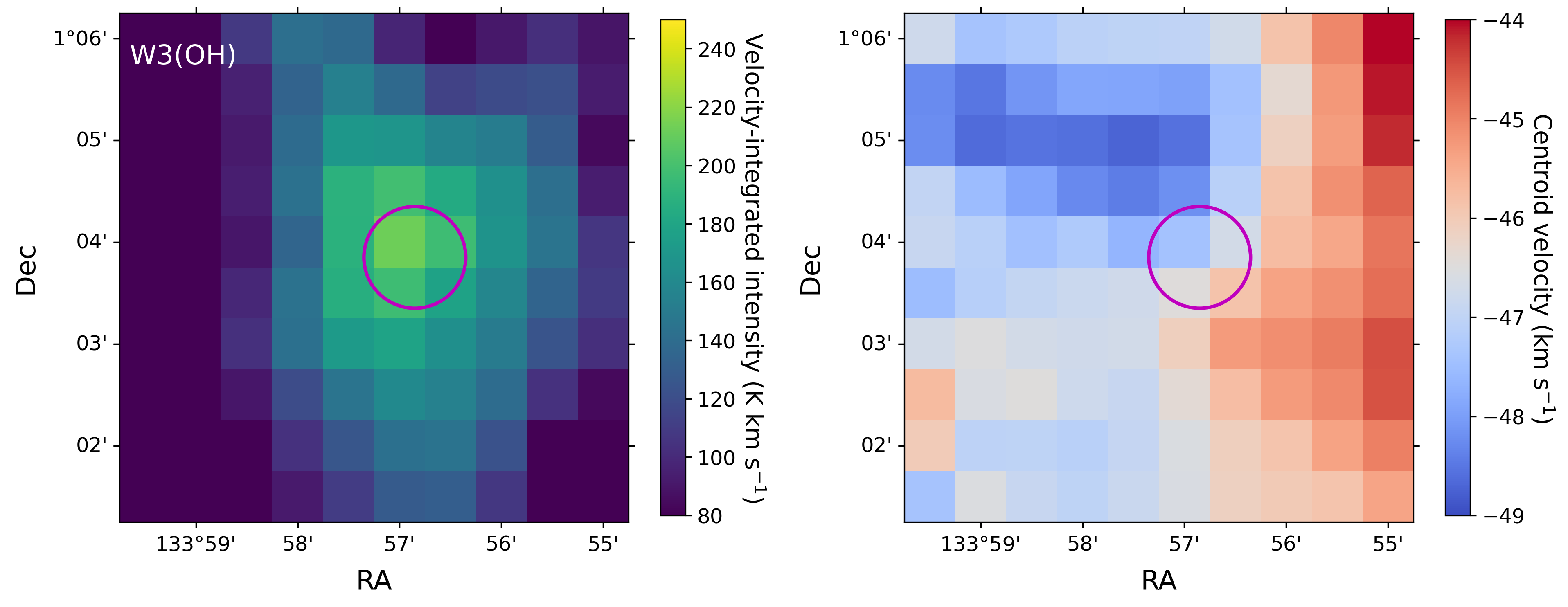}
   \caption{The distributions of the velocity-integrated intensity (left) and centroid velocity (right) 
   for the extended sources W3(OH), Sh2-140, NGC2264, and W51D.  
   These maps are derived from the On-the-fly (OTF) mapping observations of the $^{12}$CO lines 
   in the regions containing these sources. The magenta circles indicate the positions of the 
   reference sources, and their sizes are comparable to the beam size of PMO 13.7 m telescope. \label{fig:source_im}}
\end{figure*}

We further present the distributions of velocity-integrated intensity and 
centroid velocity of these extended sources in Figure \ref{fig:source_im}. 
These maps are derived from the On-the-Fly (OTF) mapping observations of $^{12}$CO spectral lines. 
The OTF mapping $^{12}$CO data for the W3(OH) and W51D sources are from the MWISP survey, 
while the data for sources Sh2-140 and NGC2264 were observed using 13.7m telescope of PMO 
to test the stability of the system's performance. 
However, the OTF mapping data for source L134 were not observed using the PMO 13.7m telescope, 
thus its maps are not included here. 
According to Figure \ref{fig:source_im}, we observe that the intensity gradually 
decreases as moving away from the position of the observed reference sources. 
Additionally, the centroid velocity around these reference sources 
shows significant variations, especially for sources W3(OH) with a velocity gradient of 1 km s$^{-1}$. 
Notably, the source W3(OH) exhibits the largest velocity shift, 
corresponding to the most pronounced variation in centroid velocity.    
In contrast, the surrounding velocity field of the NGC2264 source varies more gently, 
with velocity shifts smaller than those observed for W51D, Sh2-140, and W3(OH) sources. 
This indicates that telescope pointing errors are more likely to cause velocity shifts in these sources.
According to the annual status report of the PMO 13.7m telescope, 
the RMS error of pointing is about 5 arcsec, with peak-to-peak values up to 20 arcsec.  
For source W51D, whose CO line profile has multiple velocity components, 
the changes in centroid velocity are likely due to the varying intensities of these different velocity components. 
Therefore, its change in centroid velocity is probably not comparable to the velocity shifts of lines. 

\begin{figure*}[ht]
   \centering
   \includegraphics[width=14cm, angle=0]{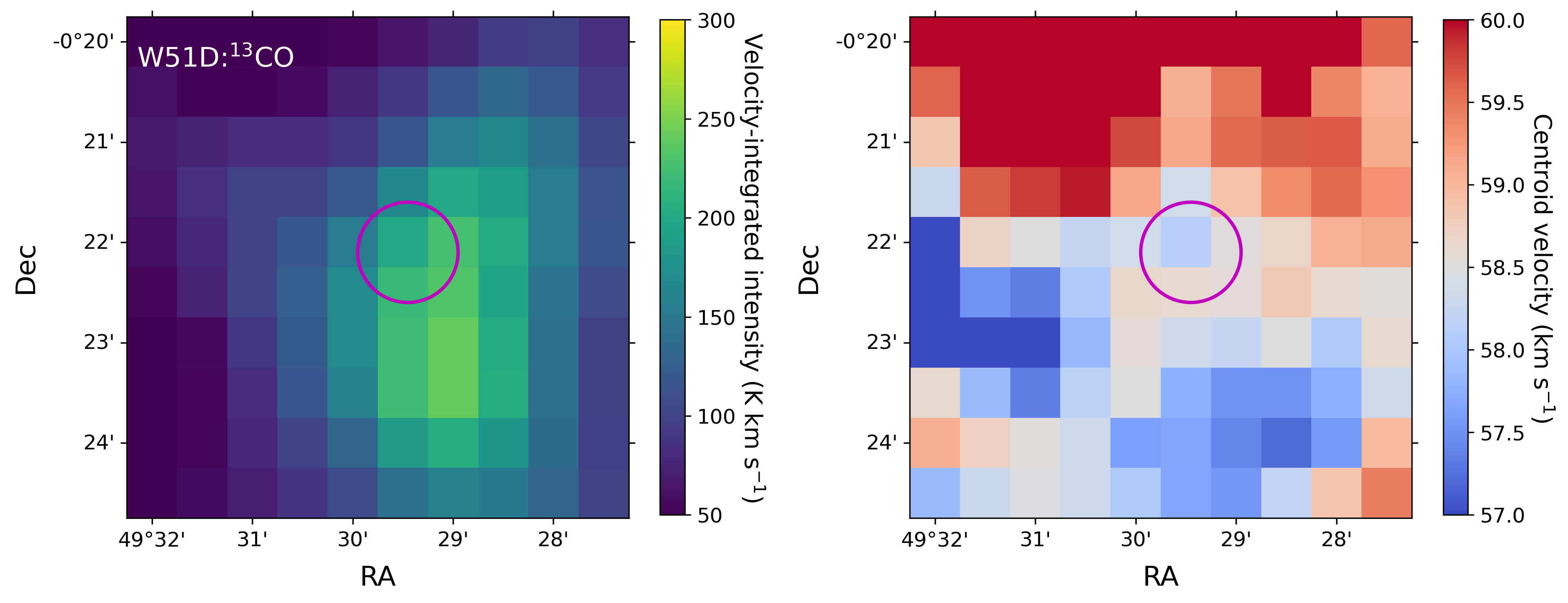}
   \includegraphics[width=14cm, angle=0]{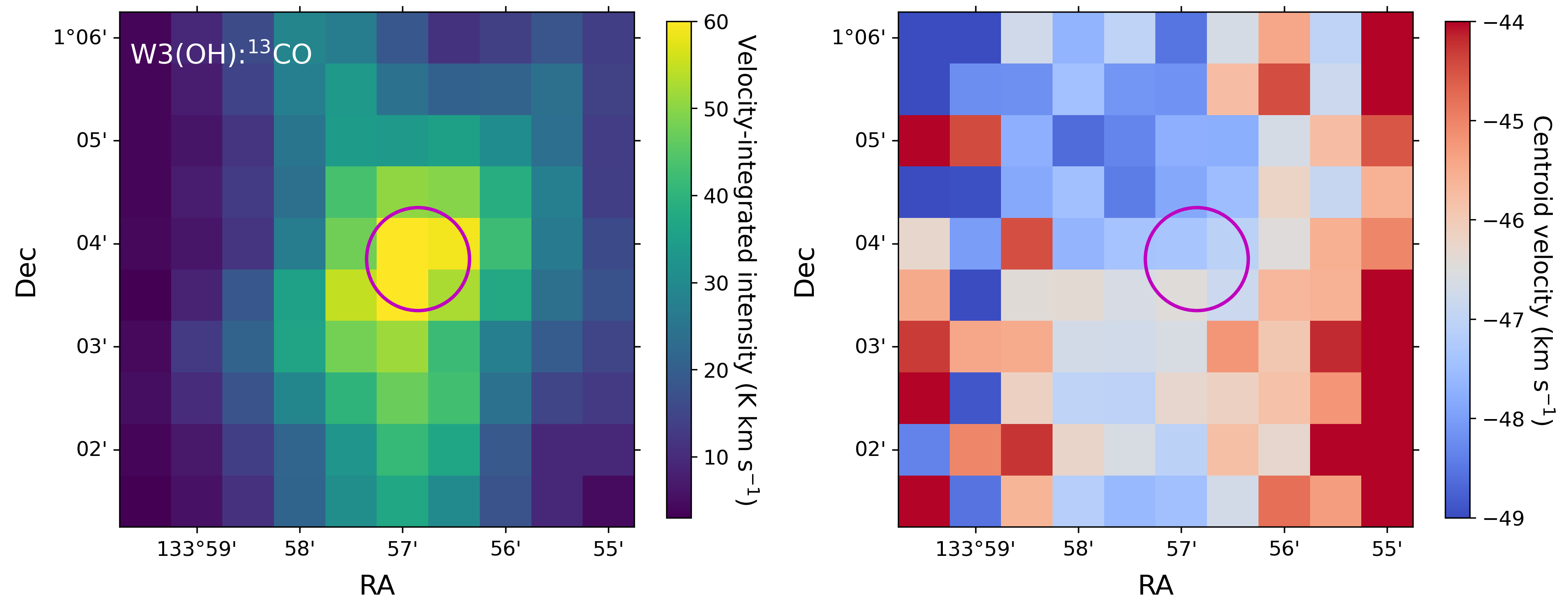}
   \caption{The distributions of the velocity-integrated intensity (left) and centroid velocity (right) 
   for the extended sources W3(OH) and W51D. 
   These maps are derived from the On-the-fly (OTF) mapping observations of the $^{13}$CO lines 
   in the regions containing these sources. The magenta circles indicate the positions of the 
   reference sources, and their sizes are comparable to the beam size of PMO 13.7 m telescope. \label{fig:source_im2}}
\end{figure*}

In addition, we also present the distributions of velocity-integrated intensity (left) and centroid velocity 
(right) derived from the OTF mapping observation of $^{13}$CO lines in Figure \ref{fig:source_im2}. 
Compared with $^{12}$CO lines, $^{13}$CO lines trace relative dense structures. 
For W3(OH) source, we find that the velocity gradient directions from $^{13}$CO lines is similar to that 
derived from $^{12}$CO lines, but the velocity gradient values are different from $^{12}$CO and $^{13}$CO lines. 
However, for source W51D, both directions and values of velocity gradients from $^{12}$CO and $^{13}$CO lines 
are different within the beam. This is consistent with that the strong correlation between $^{12}$CO and $^{13}$CO velocity shifts 
for W3(OH) source and a weaker correlation for W51D source. In addition, this also indicates the differences of 
velocity shifts from $^{12}$CO and $^{13}$CO lines are primarily due to their different velocity fields. 

\subsubsection{The distribution of velocity shifts in the Az-El space}
\begin{figure*}[ht]
   \centering
   \includegraphics[width=15cm, angle=0]{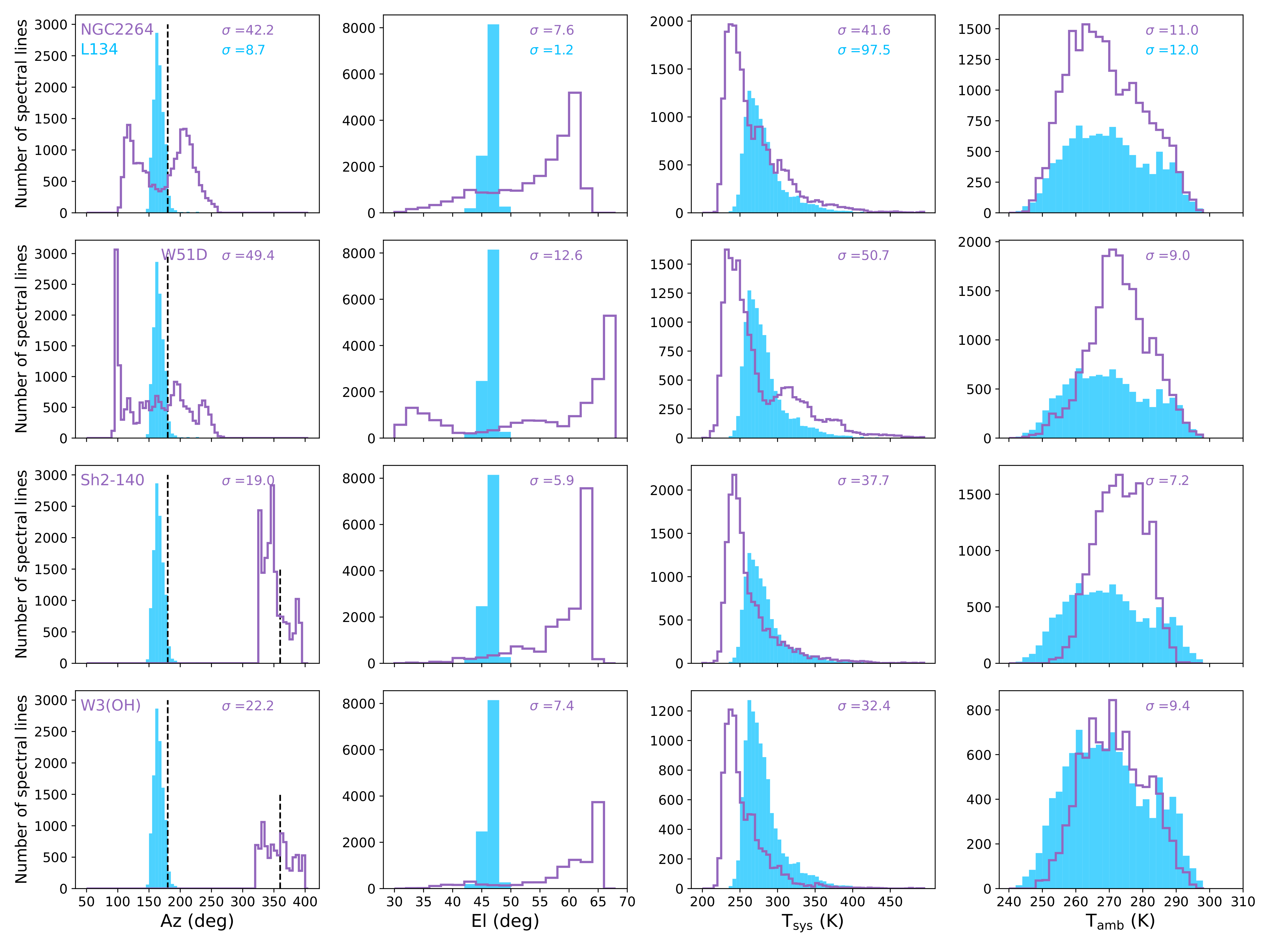}
   \caption{The distribution of the observational factors of the telescope, 
   including the Az and El coordinates, as well as the ambient temperature (T$_{\rm amb}$), 
   and the system temperature (T$_{\rm sys}$) for reference sources. 
   The blue histograms represent the factors for source L134, 
   and the purple histograms represent factors for other sources.
   The corresponding standard deviation for each factor is indicated in the corresponding panel. \label{fig:f_envir}}
\end{figure*}

We further analyze the observational factors of each reference source to 
identify the underlying causes for the increases in their velocity shifts. 
Figure \ref{fig:f_envir} illustrates the distributions of Az, El, T$_{\rm amb}$, and T$_{\rm sys}$ for each reference source. 
Among these sources, the parameters for L134, 
which experiences the lowest velocity shifts, are represented as blue histograms and 
compared with those for other sources. 
Sources L134, NGC2264, and W51D are observed in the southern sky within the horizon coordinate system, 
where the telescope's transit occurs at an Az of 180 degrees.  
For L134, the Az primarily ranges between 160 and 190 degrees, 
which are close to the transit range of the telescope. 
Its El is mostly concentrated between 45 and 48 degrees, 
and its system temperature is approximately 30 K higher than that of other sources. 
In constrast, sources NGC2264 and W51D have Az values mainly distributed between 100 and 250 degrees, 
with their El spanning from 30 to 70 degrees, indicating a wider Az-El range than L134.
On the other hand, sources W3(OH) and Sh2-140 are observed in the northern plane of 
the horizon coordinate system. When they cross the transit, the Az is either at 360 degrees or 0 degrees. 
As a result, their Az is primarily distributed between 310 degrees and 400 degrees, 
while their El mainly ranges from 50 to 65 degrees.

\begin{figure*}[ht]
   \centering
   \includegraphics[width=15cm, angle=0]{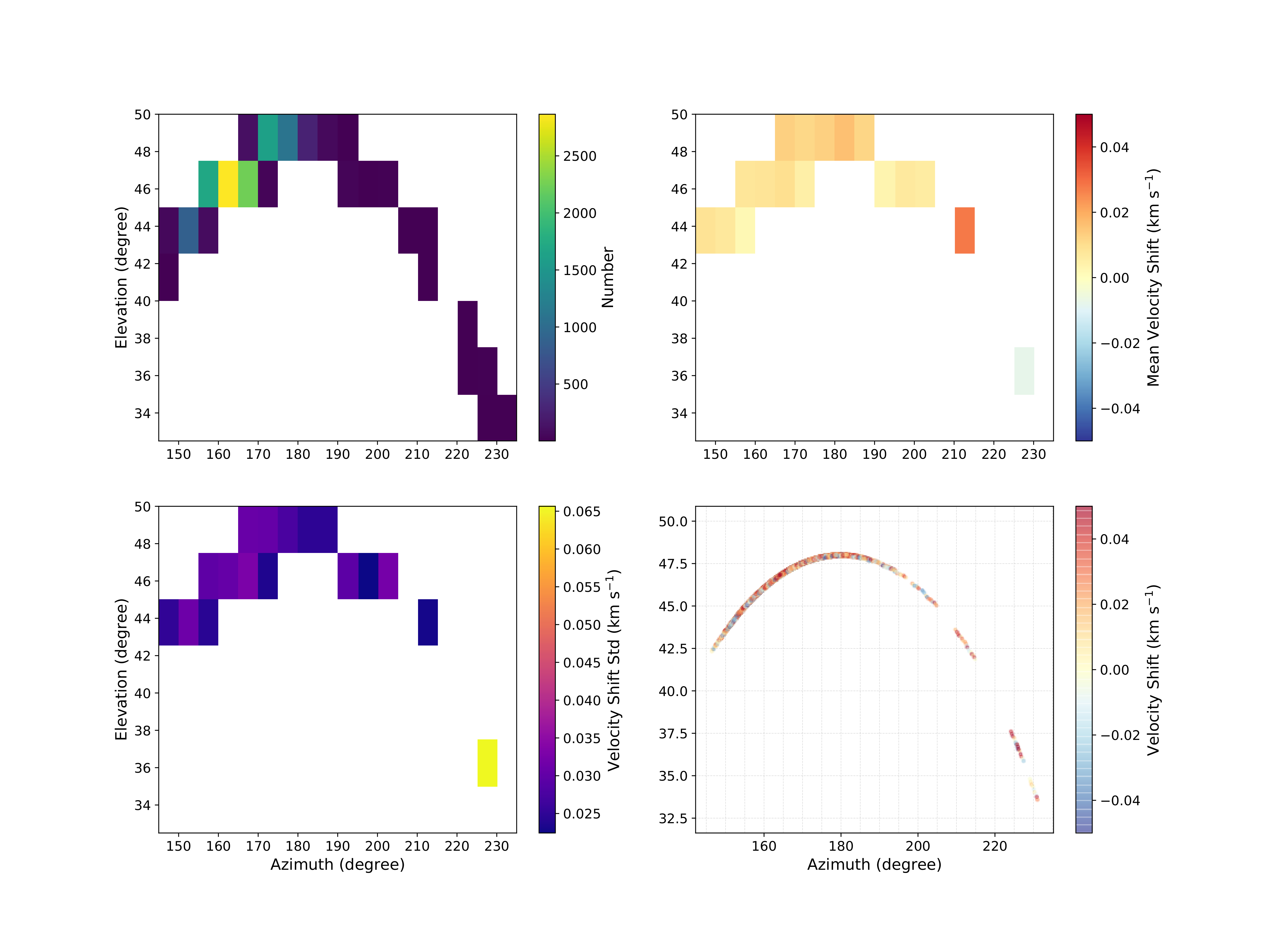}
   \caption{The distribution of the observed spectral line number (upper-left panel), 
   the mean velocity shifts (upper-right panel), the standard deviation (std) of velocity shifts (lower-left) in the 2D-bins of Az and El space for source L134. 
   Each 2D-bin has the size of Az with 5 degrees and El with 2.5 degrees. 
   The whole velocity shifts distributed in Az-El space are also presented in lower-right panel. \label{fig:L134_2D}}
\end{figure*}

\begin{figure*}[ht]
   \centering
   \includegraphics[width=15cm, angle=0]{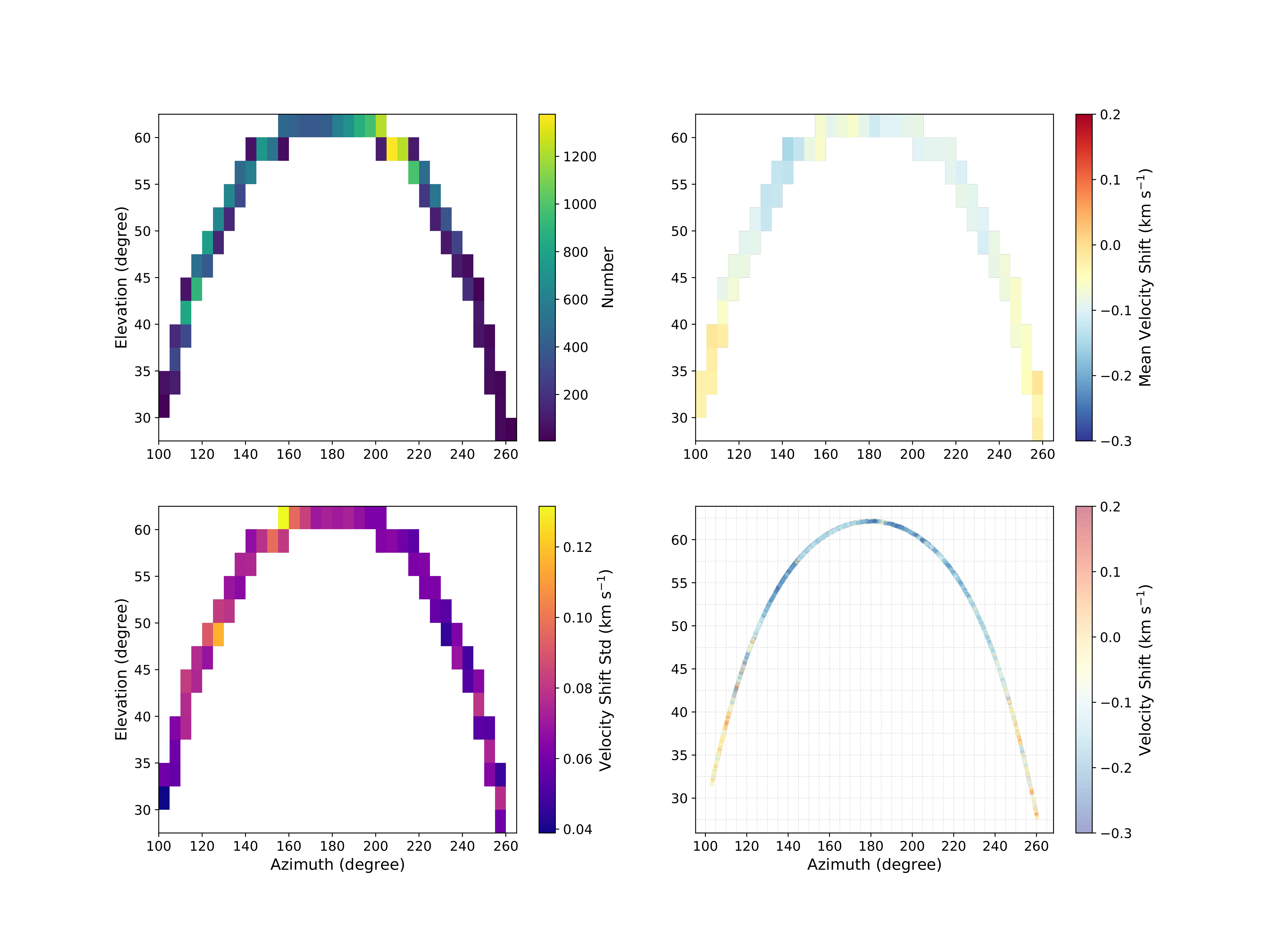}
   \caption{Same with Figure \ref{fig:L134_2D}, but for source NGC2264. \label{fig:NGC2264_2D}}
\end{figure*}

\begin{figure*}[ht]
   \centering
   \includegraphics[width=15cm, angle=0]{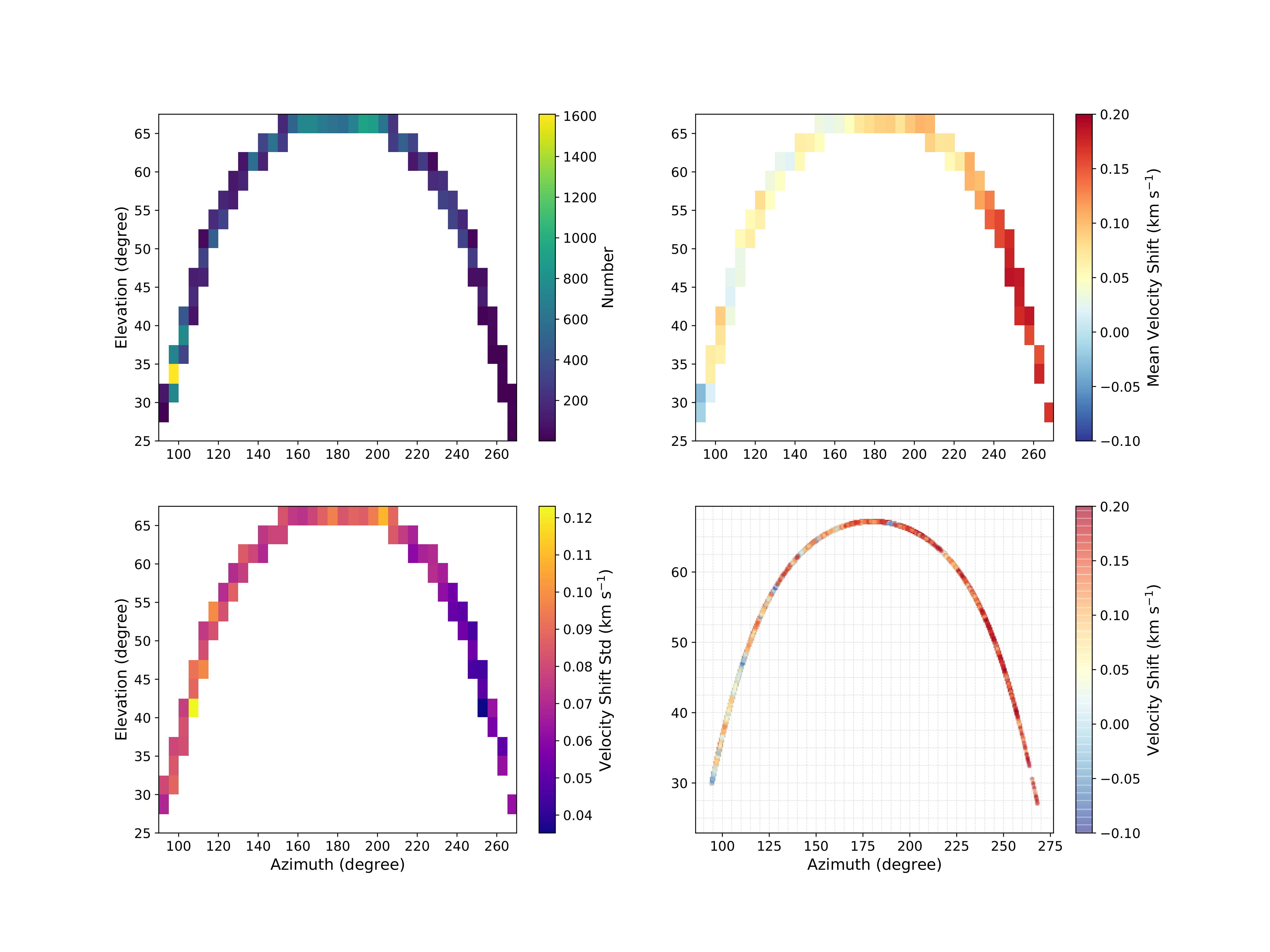}
   \caption{Same with Figure \ref{fig:L134_2D}, but for source W51D. \label{fig:W51D_2D}}
\end{figure*}

\begin{figure*}[ht]
   \centering
   \includegraphics[width=15cm, angle=0]{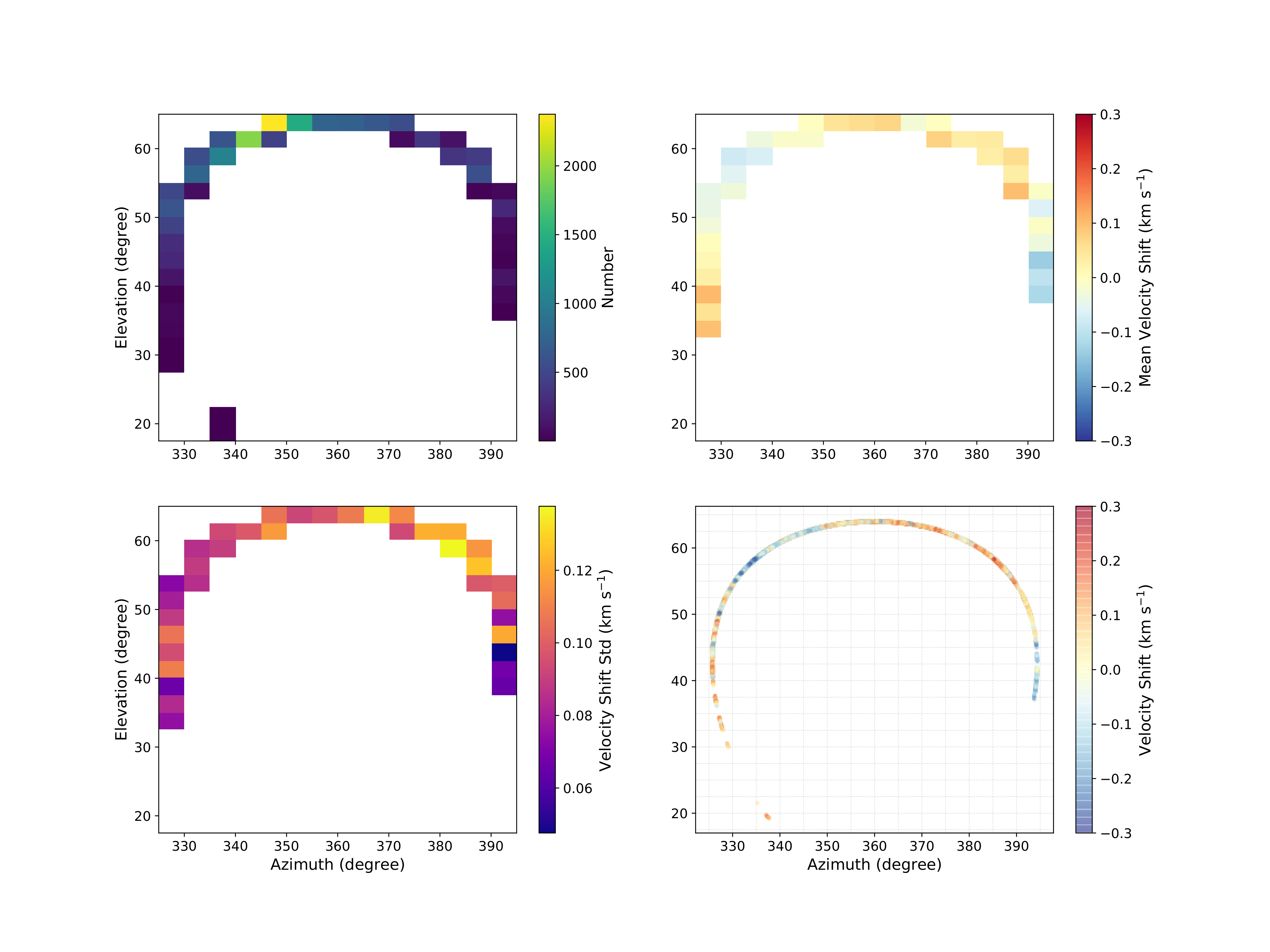}
   \caption{Same with Figure \ref{fig:L134_2D}, but for source S140. \label{fig:S140_2D}}
\end{figure*}

\begin{figure*}[ht]
   \centering
   \includegraphics[width=15cm, angle=0]{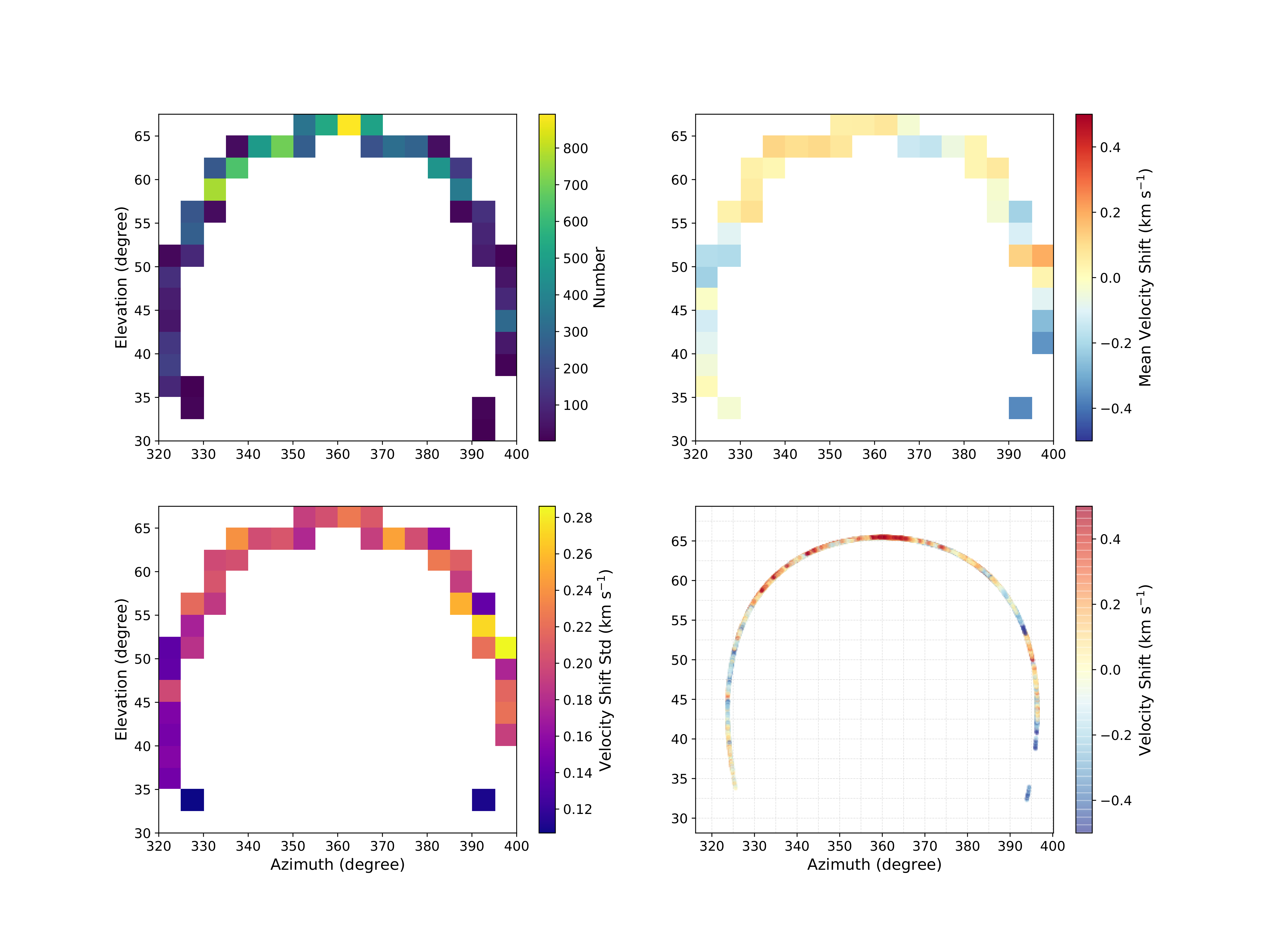}
   \caption{Same with Figure \ref{fig:L134_2D}, but for source W3(OH). \label{fig:W3OH_2D}}
\end{figure*}

In comparison to the source L134, which exhibits the smallest velocity shifts, 
the primary differences with other sources lie in the distributions of their Az and El. 
The system temperature of L134 is higher than that of other sources due to its lower elevation.
According to that, we further investigate the relation between velocity shifts and 
their corresponding Az and El.
We divide the Az-El space of sources into 2D bins, with elevation increments of 2.5$^{\circ}$ 
and azimuth increments of 5$^{\circ}$. 
Within each bin, we calculate the number of spectral lines, the mean velocity shifts, 
and the standard deviation of velocity shifts. 
The results are presented in Figures \ref{fig:L134_2D} to \ref{fig:W3OH_2D} for each source. 
Additionally, the distribution of velocity shifts in the Az-El spaces is also illustrated. 
We find that the distribution of velocity shifts in Az-El space exhibits systematic changes, 
though these changes varies for different sources.
In the southern sky of the horizon coordinate system, 
the distribution of velocity shifts for source L134 has no significant change in its Az-El range. 
In contrast, for source NGC2264, there is also a systematic change of about 0.1 km s$^{-1}$ for the El below 45$^{\circ}$ 
compared to those above 45$^{\circ}$. In addition, the standard deviation of velocity shifts in the Az range from 90$^{\circ}$ to 180$^{\circ}$ 
is approximately 0.05 km s$^{-1}$ higher than that from 180$^{\circ}$ to 270$^{\circ}$. 
For source W51D, the mean velocity shifts display a systematic change of $\sim$ 0.1 km s$^{-1}$ between 
the Az ranges of 90$^{\circ}$ to 180$^{\circ}$ and 180$^{\circ}$ to 270$^{\circ}$. 
Furthermore, the standard deviation of velocity shifts in the Az range from 90$^{\circ}$ to 180$^{\circ}$ 
is roughly 0.05 km s$^{-1}$ higher than in the range from 180$^{\circ}$ to 270$^{\circ}$.
In the northern sky, for source S140, the mean velocity shifts in the 2D bins of the Az-El space 
show fluctuations with an amplitude of approximately 0.1 km s$^{-1}$ along both Az and El. 
For source W3(OH), the mean velocity shifts also exhibit fluctuations along both Az and El, 
but with a greater amplitude of around 0.2 km s$^{-1}$.

\subsubsection{The distribution of velocity shifts on different beams}
\begin{figure*}[ht]
   \centering
   \includegraphics[width=15cm, angle=0]{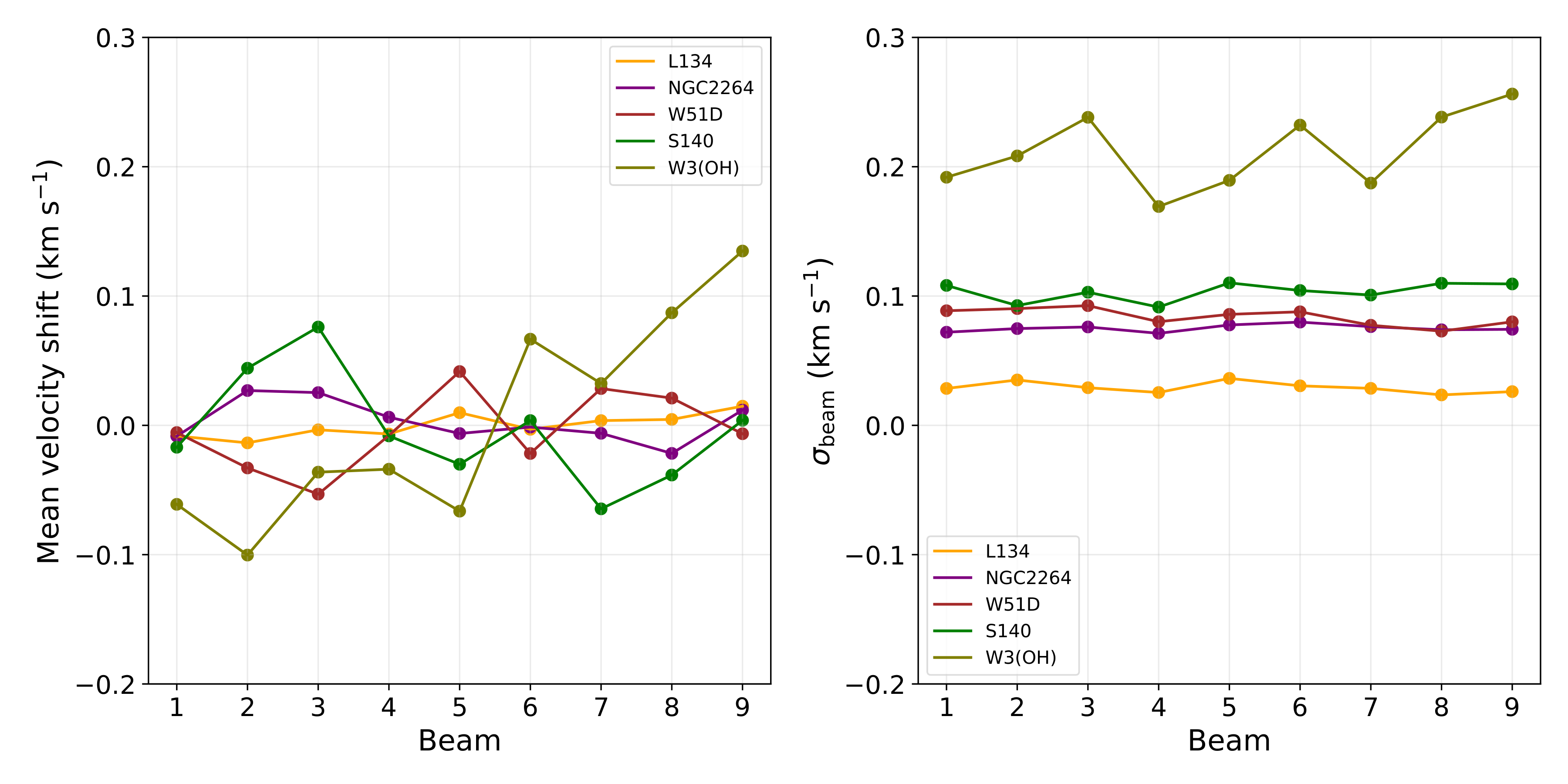}
   \caption{Mean velocity shifts and standard deviation ($\sigma_{\rm beam}$) of velocity shifts in different beams for 
   each reference source. \label{fig:f_beam}}
\end{figure*}

The spectral lines of reference sources are observed using a 3$\times$3 multibeam receiver 
in PS mode, sequentially moving from beam 1 to beam 9. 
We further investigate the performance of different beams in terms of velocity stability. 
In Figure \ref{fig:f_beam}, 
we present the mean velocity shifts and the standard deviations ($\sigma_{\rm beam}$) of $^{12}$CO velocity shifts 
across nine beams for each source. 
The changes in velocity shifts among the beams differs depending on the reference sources. 
For source L134 and NGC2264, both the mean and $\sigma_{\rm beam}$ of velocity shifts across each beam are relatively stable.
For source W51D and S140, 
the mean velocity shift between each beam varies by approximately 0.05 to 0.1 km s$^{-1}$, 
which is smaller than their $\sigma_{\rm beam}$ values. 
Their $\sigma_{\rm beam}$ values for velocity shifts in each beam are also stable and 
are close to the corresponding $\sigma_{\rm ^{12}CO}$ values for the overall velocity shifts across all beams. 
For source W3(OH), the mean velocity shifts among beams exhibits the most noticeable changes, 
with differences of about 0.05 to 0.2 km s$^{-1}$. 
However, its $\sigma_{\rm beam}$ of velocity shifts in each beam is around 0.2 km s$^{-1}$ 
varying by 0.05 km s$^{-1}$. Overall, for each source, 
the differences in mean velocity shifts between beams are smaller than their $\sigma_{\rm beam}$ values.
Therefore, there is no significant difference in velocity shifts among the nine beams.

After presenting the distribution of velocity shifts in the Az-El space, different beams, 
and the velocity fields on the extended sources, 
we find that the velocity shifts exhibit systematic changes across different Az-El ranges. 
However, both the mode and amplitude of these changes differ among the different sources. 
This variation is related to the distribution of the velocity field on these extended sources. 
Notably, the observed velocity shifts are more significant for sources that have greater velocity changes. 
This indicates that the increase of velocity shifts are mainly caused by pointing errors, 
which is more sensitive to extended sources with higher velocity gradients.

\section{Conclusions} 
We employed the cross-correlation method to measure the velocity shifts among $\sim$ 10,000 spectral lines 
of reference sources observed during the 10-year period of the MWISP survey. 
Our goal is to assess the velocity stability of CO lines from the MWISP survey and 
to identify factors that influence the long-term stability of radial velocity. 
This research is crucial for enhancing the observed velocity accuracy of the telescope facilities.  
Our main findings are as follows: 

1. The standard deviations of measured velocity shifts ($\sigma$) for six reference sources 
range from 0.03 to 0.23 km s$^{-1}$ for their $^{12}$CO spectral lines. 
In contrast, the $\sigma$ values for their $^{13}$CO spectral lines are generally smaller, 
ranging from 0.03 to 0.16 km s$^{-1}$.

2. The $\sigma$ of measured velocity shifts for sources with wider linewidths tends to be greater than 
that for sources with narrower linewidths.

3. The disparity between the overall $\sigma$ and the monthly value ($\sigma_{\rm m}$) 
increases for sources with larger overall shifts.

4. The correlation between velocity shifts of $^{12}$CO lines and those of $^{13}$CO lines 
strengthen with increasing $\sigma$. 

5. The velocity shifts show systematic variation across different Az–El ranges, 
with both the behavior and amplitude varying from source to source. 
These differences are strongly linked to the intrinsic velocity field of the extended reference sources: 
larger velocity gradients correspond to more pronounced shifts.

These findings demonstrate that MWISP achieves a high level of radial velocity stability, 
making it suitable for studying Galactic molecular gas. Additionally, this research also identifies 
that the observational sources with velocity gradients are sensitive to telescope pointing errors. 
The high signal-to-noise template spectra provided serve as valuable calibration references for future 
MWISP observations and related CO surveys.

\begin{acknowledgements}
   We are greatful for Yang Su, Fujun Du, Xin Zhou, and Xuepeng Chen for their helpful discussion.
    This research was supported by the National Natural Science Foundation of China through grant 
    12303034 $\&$ 12041305 and the Natural Science Foundation of Jiangsu Province through grant BK20231104.
    This research made use of the data from the Milky Way Imaging Scroll Painting (MWISP) project, 
    which is a multi-line survey in $^{12}$CO/$^{13}$CO/C$^{18}$O along the northern galactic plane with PMO-13.7m telescope. 
    We are grateful to all of the members of the MWISP working group, particulaly the staff members at the PMO-13.7m telescope, 
    for their long-term support. MWISP was sponsored by the National Key R\&D Program of China with grant 2023YFA1608000 $\&$ 2017YFA0402701 and 
    the CAS Key Research Program of Frontier Sciences with grant QYZDJ-SSW-SLH047.
\end{acknowledgements}

\clearpage
\appendix
%\restartappendixnumbering
\renewcommand{\thefigure}{\Alph{section}\arabic{figure}}
\renewcommand{\theHfigure}{\Alph{section}\arabic{figure}}
\renewcommand{\thefigure}{A\arabic{figure}}

\setcounter{figure}{0}

\begin{figure*}[ht]
   \centering
   \includegraphics[width=14cm, angle=0]{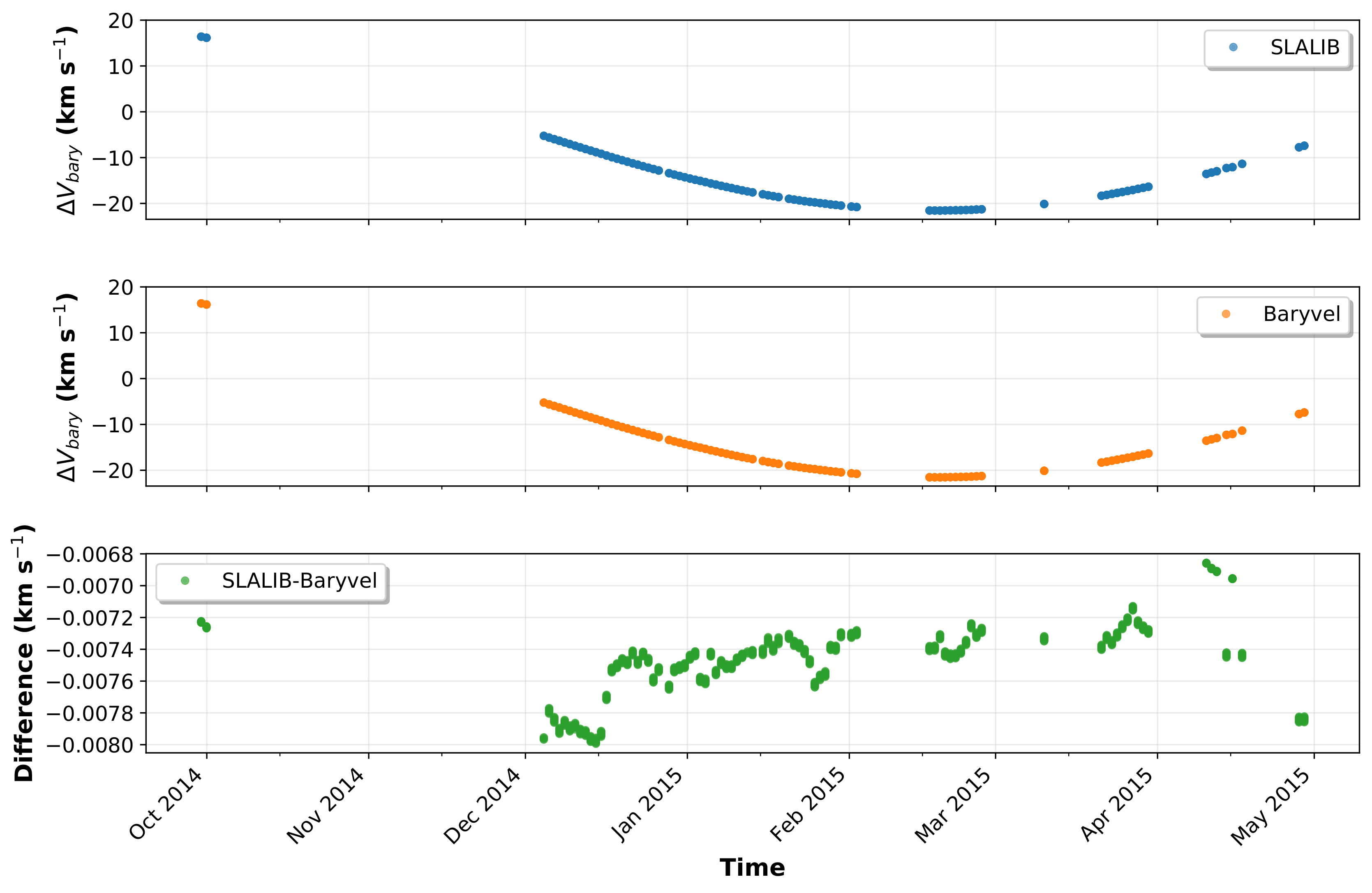}
   \caption{Upper plane: the barycentric velocities for source W3(OH) computed by the SLALIB using in the 
   MWISP CO survey. Middle plane: the barycentric velocities for source W3(OH) computed by the baryvel procedure. 
   Lower plane: the differences between the computed values from SLALIB library and baryvel procedure. \label{fig:f_vbary}}
\end{figure*}

\clearpage
\bibliography{stand_source.bib}{}
\bibliographystyle{raa}

\label{lastpage}

\end{document}